\documentstyle[12pt]{article}
\def\fcskp{\baselineskip 12pt}              % For figure captions
\newcommand{\fc}{\footnotesize \fcskp \advance\itemsep by -6pt}
\def\half{{\textstyle{1\over2}}}
\def\frac#1#2{{\textstyle{#1\over#2}}}
\def\del{\partial}
\def\Tr{\mathop{\rm Tr}\nolimits}
\def\bar#1{\overline{#1}}
\def\vev#1{\langle #1 \rangle}
\def\g#1{\gamma_{#1}}
\def\pl{(1\!+\!\g5)}
\def\pr{(1\!-\!\g5)}
\def\cm{{\cal M}}
\def\co{{\cal O}}
\def\cl{{\cal L}}
\def\cs{{\cal S}}
\def\ca{{\cal A}}
\def\vac{{\rm vac}}
\def\YtabC{\vbox{\hbox{$\sqr\sqr\thinspace$}\nointerlineskip
        \kern-.3pt\hbox{$\sqr\sqr$}}}
\def\YtabI{\lower 1ex\hbox{\vbox{\hbox{$\sqr\sqr\sqr\sqr\thinspace$}
    \nointerlineskip\kern-.3pt\hbox{$\sqr\sqr$}
    \nointerlineskip\kern-.3pt\hbox{$\sqr\sqr$}}}}

\begin{document}
%----------------------------------------------------------------------%
\begin{titlepage}
 \null
 \begin{center}
 \makebox[\textwidth][r]{CEBAF-TH-92-12}
 \par\vspace{1.25in} %%%
  {\Large
	QUENCHED CHIRAL LOGARITHMS}
  \par
 \vskip 2.0em
 {\large
  \begin{tabular}[t]{c}
	Stephen R. Sharpe \footnotemark\\[1.em]
	\em Continuous Electron Beam Accelerator Facility \\
	\em Newport News, VA 23606 \\[1.em]
  \end{tabular}}
 \par \vskip 3.0em
 {\large\bf Abstract}
\end{center}
\quotation
I develop a diagrammatic method for calculating chiral logarithms
in the quenched approximation.
While not rigorous, the method is based on physically reasonable
assumptions, which can be tested by numerical simulations.
The main results are that, at leading order in the chiral expansion,
(a) there are no chiral logarithms in quenched $f_\pi$, for $m_u=m_d$;
(b) the chiral logarithms in $B_K$ and related kaon B-parameters are,
for $m_d=m_s$,
the same in the quenched approximation as in the full theory;
(c) for $m_\pi$ and the condensate, there are extra chiral logarithms
due to loops containing the $\eta'$, which lead to a peculiar non-analytic
dependence of these quantities on the bare quark mass.
Following the work of Gasser and Leutwyler, I discuss how
there is a predictable finite volume dependence
associated with each chiral logarithm.
I compare the resulting predictions with numerical results:
for most quantities the expected volume dependence is smaller than the
errors, but for $B_V$ and $B_A$ there is an observed dependence which is
consistent with the predictions.

\endquotation

%\vspace{.5in}

\footnotetext{Permanent address:
Physics Department, FM-15, University of Washington, Seattle, WA 98195}
\vfill
\mbox{April 1992}
\end{titlepage}

% beginning of body
\section{INTRODUCTION}
\label{sintro}
%%% sec1.tex
%% files: sec1.tex
%\section{INTRODUCTION}
%\label{sintro}

The approximate chiral symmetry of QCD provides strong constraints
on the form of amplitudes involving pseudo-Goldstone bosons.
A classic example is that the pion scattering amplitudes
at threshold are related to the pion decay constant \cite{weinberg}
\begin{equation}
T_I^{\pi\pi}(s\!=\!4m_\pi^2,t\!=\!u\!=\!0)
= c_I {m_\pi^2 \over f_\pi^2} + O(m_\pi^4) \ .
\end{equation}
The $c_I$ are known numerical constants which depend on the isospin $I$
of the pion pair.
The corrections of $O(m_\pi^4)$ fall into two classes:\footnote{%
%%%%%%%%%%%%%%%%%%%%%%%%%%%%%%%%%%%%%%%%%%%%%%%%%%%%%%%%%%%%%%%%%%%%%%%%
Here and in the following I often use ``pion'' to generically to refer
to all of the pseudo-Goldstone bosons, i.e. the pions, kaons and eta's.
In the present example,
$O(m_\pi^4)$ includes the possibility of $O(m_\pi^2 m_K^2)$.}
%%%%%%%%%%%%%%%%%%%%%%%%%%%%%%%%%%%%%%%%%%%%%%%%%%%%%%%%%%%%%%%%%%%%%%%%
those that are analytic in $m_\pi^2$, coming from non-leading terms
in the chiral Lagrangian \cite{gassleut};
and the non-analytic ``chiral logarithms'' \footnote{%
%%%%%%%%%%%%%%%%%%%%%%%%%%%%%%%%%%%%%%%%%%%%%%%%%%%%%%%%%%%%%%%%%%%%%%%%
For baryon properties some of the non-analytic terms are not logarithms,
but are proportional to $(m_\pi^2)^{3/2}$ \cite{langackerpagels}.
I do not consider baryon properties here.}
%%%%%%%%%%%%%%%%%%%%%%%%%%%%%%%%%%%%%%%%%%%%%%%%%%%%%%%%%%%%%%%%%%%%%%%%
proportional to $m_\pi^4 \ln(m_\pi)$. These logarithms
come from infrared divergences in pion loops
\cite{lipagels,langackerpagels}.
They have the important property that they are proportional to the
lowest order term in the chiral expansion. For example,
in the limit $m_u\!=\!m_d\!=\!0$, $f_\pi$ has the expansion
\cite{langackerpagels,gassleut}
\begin{equation}
\label{fpireallogeqn}
f_\pi = f (1 - \frac12 L(m_K) + c m_K^2)\ .
\end{equation}
where the chiral logarithm is
\begin{equation}
\label{chirallogdef}
L(m) = \left({m \over 4\pi f}\right)^2 \ln\left({m^2\over\Lambda^2}\right)\ .
\end{equation}
(I use the normalization such that $f_\pi=93$ MeV.)
Chiral symmetry alone does not determine $f$, $\Lambda$ or $c$
(in fact $c$ can be absorbed into $\Lambda$),
but does fix the coefficient of the chiral log $L$.

In the continuum, chiral logs are useful as a guide to the size of
the corrections to the leading order chiral behavior.
If the logs are large when one takes a reasonable scale $\Lambda\sim m_\rho$,
then there is reason to worry about the reliability of
leading order chiral predictions.
This approach has been taken for example in Refs.
\cite{langackerpagels} and \cite{sonoda}.

On the lattice, chiral logs are potentially of greater importance,
because they allow one to test whether the contribution of pion loops
is being correctly described.
These loops give rise not only to chiral logs,
but also to the pion cloud surrounding hadrons.
This cloud is an important part of the structure of hadrons,
affecting charge radii and other form factors.
The obvious way to search for the chiral log,
by studying the dependence of amplitudes on $m_\pi^2$ or $m_K^2$,
is not very useful,
for it is difficult to disentangle a logarithm from a power series.
It is better to study the volume dependence of the amplitude.
As explained by Gasser and Leutwyler \cite{condfinitev},
associated with each chiral log is a known volume dependent correction.
For the particular case of the masses, the correction can also be
obtained from a general formula given by L\"uscher \cite{luscherexp}.

Chiral logarithmic corrections have been calculated for a number of a
quantities in QCD. Those that are of interest here
are the pseudo-Goldstone boson masses and decay constants,
the condensate, and $B_K$.
I have extended these calculations to $B_V$ and $B_A$.
These results can be converted into predictions of finite volume dependence
for lattice simulations of full QCD, and I give some examples below.
At present, however, most lattice calculations of these quantities use
the so-called ``quenched'' approximation,
in which internal quark loops are not included.
The lack of internal quark loops means that many, though not all,
of the pion loops present in full QCD are absent in the quenched approximation.
The main focus of this article is to understand how to adapt the
calculation of chiral logarithms in full QCD to the quenched approximation.
This is done here for the quantities just mentioned,
but the method can be extended to many other quantities.
These results allow one to predict the finite volume dependence of
quenched matrix elements.

In addition to their diagnostic power, chiral logs provide an
estimate of the size of the error introduced by the quenched approximation.
In most quantities the chiral logs have
a different coefficient in quenched and full theories.
For example, it is shown below that there are no leading order
chiral logs in $f_\pi$ in the quenched approximation,\footnote{%
%%%%%%%%%%%%%%%%%%%%%%%%%%%%%%%%%%%%%%%%%%%%%%%%%%%%%%%%%%%%%%%%%%%%%%%%%
Strictly speaking this is only true if $m_u=m_d$, as discussed below.}
%%%%%%%%%%%%%%%%%%%%%%%%%%%%%%%%%%%%%%%%%%%%%%%%%%%%%%%%%%%%%%%%%%%%%%%%%
in contrast to the result for full QCD, Eq. (\ref{fpireallogeqn}).
Taking $\Lambda=m_\rho$, the chiral log in full QCD gives a 6\% contribution
to $f_\pi$.
This suggests that the absence of chiral logs in the quenched approximation
should have an effect on $f_\pi$ of this size.
This is a suggestion, and not a calculation, because the
quantities $f$ and $c$ need not be the same in full and quenched QCD.
Nevertheless, even a rough guide is useful since
we have so few tools for studying the effect of quenching.

The idea of calculating chiral logarithms in the quenched approximation
was first suggested by Morel \cite{morel}, who was inspired by numerical
evidence of non-analytic terms in $m_\pi^2/m_q$ (Ref. \cite{saclay}).
In order to study the quenched approximation,
he introduced scalar-quark ghost fields which, upon functional integration,
produce a determinant which cancels that from the integration over quarks.
He showed how to calculate chiral logs in the combined strong coupling,
large dimension limit, both for color gauge groups $SU(2)$ and $SU(3)$.
In Ref. \cite{sschlog}
I corrected errors in Morel's result, finding that, for three colors,
there were {\em no} quenched chiral logs in $f_\pi$ or $m_\pi$ in
the large $d$, strong coupling limit.\footnote{%
%%%%%%%%%%%%%%%%%%%%%%%%%%%%%%%%%%%%%%%%%%%%%%%%%%%%%%%%%%%%%%%%%%%%%%%%%%%
The reader may be confused by two points upon consulting Ref. \cite{sschlog}.
(1) A class of diagrams involving $\eta'$ loops,
which are discussed at length in this article,
were overlooked in Ref. \cite{sschlog}.
This does not affect main calculation of that reference, because these diagrams
are not present in the large $d$, strong coupling limit \cite{morel}.
It does, however, mean that the conclusion drawn in that paper,
namely that there are no leading order chiral logs in quenched
$f_\pi$ and $m_\pi$ in general, is false.
This error is rectified here.
(2) There {\em are} chiral logs in $f_\pi$ and $m_\pi$ for color $SU(2)$,
even in the large $d$, strong coupling limit.
These may explain the numerical results of Ref. \cite{saclay}.
$SU(2)$ is special, however, since there are Goldstone baryons as well
as Goldstone pions. In this paper I consider only three or more colors.}
%%%%%%%%%%%%%%%%%%%%%%%%%%%%%%%%%%%%%%%%%%%%%%%%%%%%%%%%%%%%%%%%%%%%%%%%%%%
More importantly, I showed how the presence or absence
of chiral logs could be understood by studying quark diagrams.
It is the extension of this ``diagrammatic method'' that I use in this paper.

As will become clear in the following,
this method is neither rigorous nor systematic.
The lack of rigor is common to all calculations of quenched chiral logs.
Such calculations use chiral perturbation theory,
the validity of which is predicated on the existence
of a well defined underlying field theory.
Quenched QCD is not a well defined field theory.
In particular, quenched amplitudes are neither unitary nor analytic.
Nevertheless, as I discuss in Sections \ref{schiral} and \ref{sfpi},
it is reasonable to assume both that chiral logs are present,
and that they can be calculated using an adaptation of chiral
perturbation theory.

The diagrammatic method is also not systematic.
It amounts to a set of rules for discarding
some of the chiral perturbation theory diagrams,
and for changing the masses of the pions that appear in some of the loops
of the diagrams that are kept.
These rules are developed in an {\em ad hoc} way, quantity by quantity.
I do not know how to write them down in general.
Bernard and Golterman \cite{bernardgolterman} have
recently outlined a more systematic approach in which chiral perturbation
theory is extended to include states containing Morel's scalar ghosts.
Their method appears to give the same answers for the quantities that I
calculate here.
I prefer to use the diagrammatic method, however,
because it displays the underlying physics very clearly.
For more complicated calculations Bernard and Golterman's method is
likely to be superior.

A difficulty with calculations of quenched chiral logarithms
stems from the fact that one must treat the flavor singlet $\eta'$
as a pseudo-Goldstone boson. This is in contrast to full QCD,
where the $\eta'$ is not a pseudo-Goldstone boson because of the
axial anomaly.
It turns out the contributions of $\eta'$ loops do not satisfy the usual power
counting rules of chiral perturbation theory, in which each new loop
brings with it a power of $m_\pi^2$.
Thus corrections proportional to powers of $\ln(m_\pi)$
are obtained from diagrams with any number of loops.
This is a potentially disastrous, for it might undermine the application
of quenched chiral perturbation theory to all quantities.
As a first attempt to understand the issue,
I have summed the leading logarithms,
and find that they can be absorbed into a change of
definition in the quark mass.
The only observable effects are that $m_\pi$, $\vev{\bar\psi\psi}$,
and related quantities should have a weak finite volume dependence.
More study of this issue is required.

Fortunately, it turns out that there are a class of ``safe'' quantities
which are insensitive, at leading order, to $\eta'$ loops.
Examples are the decay constants and kaon $B-$parameters
for degenerate quarks.
I concentrate on these quantities for most of this paper.

Bernard and Golterman have suggested a more ambitious approach
\cite{bernardgolterman}.
They point out that, although $\eta'$ loops are not parametrically
of higher order in chiral perturbation theory, they are non-leading
in a combined chiral and large $N_c$ expansion.
Furthermore, their contribution is numerically small.
Thus, from a practical point of view, it may make sense to do a loop
expansion, and they have begun such a program.

The work presented here was inspired by numerical results showing volume
dependence in $B_V$ and $B_A$ \cite{sharpelat89}.
It turns out that these results are described reasonably well
by predictions given here.
Present calculations of other quantities are not yet accurate enough
to test the predictions.

The outline of this paper is as follows.
In the next Section I recall some useful facts about the chiral Lagrangian,
and discuss its extension to the quenched approximation.
In Section \ref{sfpi}, I introduce the method of calculating chiral logs
in the quenched approximation using the example of decay constants.
Section \ref{sbk} extends this method to $B_K$ and the related parameters
$B_V$ and $B_A$.
In Section \ref{seta'} I discuss the contribution of $\eta'$ loops,
and in Section \ref{sfinite} I explain how to predict the
leading finite volume dependence from the results for chiral logs.
These predictions are confronted with numerical data in Section \ref{snumer}.
Section \ref{sconcl} contains some conclusions.

Brief discussions of some of the results of this paper have been given
previously in Refs. \cite{bkprl,sharpelat90,sharpelat89,book}.

\section{CHIRAL LAGRANGIAN}
\label{schiral}
%%% sec2.tex
% sec2.tex
%\section{CHIRAL LAGRANGIAN}
%\label{schiral}

Chiral logarithms in full QCD
are most simply calculated using the chiral Lagrangian.
In this section I first collect some standard results,
and then explain my assumptions for how these are modified
in the quenched approximation.
I use the notation of Ref. \cite{georgi}, and work in Euclidean space.
The Lagrangian is
\begin{equation}
\label{chirall}
 \cl =
 \frac14 f^2 \Tr\left(\del_\nu \Sigma \del_\nu \Sigma^{\dag}\right)
 -\frac12 f^2 \mu \Tr\left(M (\Sigma+\Sigma^{\dag})\right)
 +O(p^4,Mp^2, M^2) \ ,
\end{equation}
where $\Sigma = \exp({2 i \pi_a T_a /f})$, the $\pi_a$ being
the pseudo-Goldstone bosons fields.
The group generators $T_a$ are normalized such that
$\Tr(T_a T_b)=\frac12\delta_{ab}$.
$M$ is the quark mass matrix, in terms of which the pion mass matrix is
$(m_\pi^2)_{ab} = 4 \mu \Tr(M T_a T_b)$.
The form of $\cl$ is identical for any number of flavors.
Flavor dependence enters only through the number of pions
($N^2\!-\!1$ for flavor $SU(N)$), and through the constants $f$ and $\mu$.
%For three flavors, and in the chiral limit, $f_\pi=f$.

In QCD the mass matrix is ${\rm diag}(m_u, m_d, m_s)$.
I will assume that $m_u=m_d$, which is a good approximation for
the amplitudes I discuss, as the terms that I drop are suppressed by
$(m_u-m_d)/m_s$ compared to those that I keep.
%The fact that $m_u\ne m_d$ causes a small mixing between the $\pi_0$ and
%the $\eta$, and slightly breaks the degeneracy of the kaons.
In this isospin symmetric limit the flavor off-diagonal eigenstates
are the $\pi^\pm$, with $m_\pi^2= 2 \mu m_d$,
and the kaons $K^\pm,K_0,\overline{K}_0$ with $m_K^2=\mu (m_s +m_d)$,
while the flavor diagonal eigenstates are the $\pi^0$ ($T_a=T_3$),
which is degenerate with the $\pi^\pm$,
and the $\eta$ ($T_a=T_8$), with $m_\eta^2=\frac13 \mu (4m_s + 2m_d)$.

The left- and right-handed quark currents can be written in terms of
$\Sigma$ fields
\begin{eqnarray}
\label{lhand}
L_{\mu a} = \overline{q} T_a \gamma_\mu \frac12 \pl q
&\longrightarrow& \frac12 f^2 \Tr( T_a \Sigma \del_\mu \Sigma^{\dag}) \ ,\\
\label{rhand}
R_{\mu a} = \overline{q} T_a \gamma_\mu \frac12 \pr q
&\longrightarrow& \frac12 f^2 \Tr( T_a \Sigma^{\dag} \del_\mu \Sigma) \ .
\end{eqnarray}
The Euclidean space Dirac matrices are hermitian and satisfy
$\{\gamma_\mu,\gamma_\nu\}=\delta_{\mu\nu}$.
The expansion of $L_\mu$ in terms of the pion field is
\begin{equation}
\label{leftcurrent}
L_{\mu a} = - if \left(\Tr(T_a \del_\mu\pi)
		     -{i\over f} \Tr(T_a[\del_\mu\pi,\pi])
		     -{2\over 3f^2} \Tr(T_a[\pi[\pi,\del_\mu\pi]])
		     + O(\pi^4) \right) \ .
\end{equation}
$R_\mu$ has the same expansion except that $\pi\to-\pi$.
The pion decay constant is defined by
\begin{equation}
\label{fpidef}
\vev{0|\bar{q} T_a \gamma_\mu \gamma_5 q|\pi_b}
= \vev{0|L_{\mu a}\!-\!R_{\mu a}|\pi_b}
= p_\mu f_\pi \ .
\end{equation}
Comparing with Eq. (\ref{leftcurrent}) we see that $f_\pi=f$ at lowest order
in chiral perturbation theory.
%%%%%%%%%%%%%%%%%%%%%%%%%%%%%%%%%%%%%%%%%%%%%%%%%%%%%%%%%%%%%%%%
%%%
% I am reading the matrix element as the destruction of a pion.
% I assume \del_\mu gives -i p_\mu, where p_\mu is the incoming momentum
% (I'm looking at the a(k) coefficient in BJD(II) 12.7)
% Since the coefficient of the destruction operator is \exp(-ip_4t)
% while the actual correlator falls like \exp(-Mt), p_4=-iM on shell.
%%%
%%%%%%%%%%%%%%%%%%%%%%%%%%%%%%%%%%%%%%%%%%%%%%%%%%%%%%%%%%%%%%%%
%For a pion at rest, the matrix element in Eq. (\ref{fpidef})
%equals $-im_\pi\delta_{\mu4}$, and is thus imaginary.
%This appears to contradict the lattice convention in which $f_\pi$ is real.
%There is no disagreement, however, because the lattice definition does
%not take into account the fact that the pseudoscalar two point function
%is negative, due to the sign due to Fermi statistics.
%Thus, in the proper lattice definition, $f_\pi$ is imaginary.
%%%%%%%%%%%%%%%%%%%%%%%%%%%%%%%%%%%%%%%%%%%%%%%%%%%%%%%%%%%%%%%%

The chiral Lagrangian can be extended to include the flavor singlet
pseudoscalar meson \cite{gassleut}.
Throughout this paper I refer to this meson as the $\eta'$,
despite the fact that the physical $\eta'$ is not exactly a flavor singlet.
I use the label ``pion'' to refer to the non-singlet
pseudo-Goldstone bosons, not including the $\eta'$.
To include the $\eta'$ in the chiral Lagrangian
one begins by defining a $U(N)$ matrix
\begin{equation}
U=\Sigma e^{i\phi_0}\ ; \quad
\phi_0 = {2\eta' / (\sqrt{2N}f)} \ ,
\end{equation}
in terms of which the extended Lagrangian can be written
\begin{eqnarray}
\nonumber
 \cl &=&
+\frac14 f^2 V_1(\eta')\Tr\big[(\del_\nu U \del_\nu U^{\dag}\big]
-\frac12 f^2 V_2(\eta')\mu \Tr\big[M U\big]
-\frac12 f^2 V_2(\eta')^*\mu \Tr\big[M U^{\dag}\big] \\
\label{chiralleta}
&&\quad + V_0(\eta') + V_5(\eta')\del_\nu \eta' \del_\nu\eta' \ .
\end{eqnarray}
(The definitions of the potentials differs slightly from those of Ref.
\cite{gassleut}.)
The potential are arbitrary aside from the following conditions:
$V_0$, $V_1$, and $V_5$ are real, and even functions of their arguments;
$V_2$ satisfies $V_2(\eta')^* = V_2(-\eta')$;
and $V_1(0)=V_2(0)=1$.
The arbitrariness implies
that there are very few constraints on the couplings of the $\eta'$.
%The $\eta'$ has a mass in the chiral limit proportional to $V_0''(0)$.

This Lagrangian simplifies when $N_c$,
the number of colors, is taken to infinity.
This limit is of interest here because it is similar
to the quenched approximation.
{}From the analysis of Ref. \cite{gassleut}, it can be seen that,
when $N_c\to\infty$, one can set $V_1=V_2=1$, and $V_0=V_5=0$.
($V_0$ does contribute a constant term proportional to $N_c^2$, but
this only affects the vacuum energy.)
In other words, the Lagrangian has the same form as
the original chiral Lagrangian, Eq. (\ref{chirall}),
except that $\Sigma$ is replaced by $U$.
Dependence on $N_c$ enters only through the decay constant,
which is proportional to $\sqrt{N_c}$.
The symmetry group has thus enlarged to $U(N)_L\times U(N)_R$,
with the $\eta'$ being the additional pseudo-Goldstone boson.
The physical reason for this simplification
is that the terms in Eq. (\ref{chiralleta})
which differentiate the $\eta'$ from the pions involve intermediate gluons,
and are suppressed by powers of $1/N_c$.

Expanding the large $N_c$ Lagrangian in terms of $\Sigma$ and $\eta'$ one finds
\begin{eqnarray}
\label{chirallnc}
 \cl({N_c\to\infty}) &=& \cl_0 + \cl_m \ ,\\
\cl_0 &=&
\frac14 f^2 \left( \Tr\big[\del_\mu \Sigma \del_\mu \Sigma^{\dag}\big] +
                     {2\over f^2} \del_\mu \eta' \del_\mu\eta' \right) \ ,\\
\label{massterm}
\cl_m &=& - \frac12 f^2  \mu
\left( \cos(\phi_0) \Tr\big[M (\Sigma\!+\!\Sigma^{\dag})\big] +
 i \sin(\phi_0) \Tr\big[M (\Sigma\!-\!\Sigma^{\dag})\big] \right)
\ .
\end{eqnarray}
Three features of this result are important in the following.
\begin{itemize}
\item
The $\eta'$ couples to pions through the mass term $\cl_m$,
{\em but does not couple through the derivative term} $\cl_0$.
This is not true at finite $N_c$ because the potential $V_1$
will contain even powers of the $\eta'$ field.
\item
The masses of the pseudo-Goldstone pions differ in
general from those at $N_c=3$.
Diagonalizing the mass matrix, one finds that the $\eta'$,
$\eta$ and $\pi^0$ mix.
The eigenstates have flavor compositions $\bar s s$,
$\bar d d$ and $\bar u u$,
with masses $2 \mu m_s$, $2\mu m_d$ and $2\mu m_u$ respectively.
For $m_d=m_u$ one can use the usual $\pi^0$ and $\eta$
linear combinations, since they are degenerate.
Only if all three quarks are degenerate, however,
is the flavor singlet $\eta'$ an eigenstate.
\item
The non-singlet chiral currents that follow from the Lagrangian
are the same as those of Eqs. (\ref{lhand}) and (\ref{rhand}),
except that $\Sigma$ is replaced by $U$.
It is simple to see, however, that the $\eta'$ drops out of the expressions,
so that the currents are in fact unchanged.
This can be seen explicitly in the expansion of $L_{\mu a}$,
Eq. (\ref{leftcurrent}).
The first term vanishes for the $\eta'$ because of the flavor trace,
and all subsequent terms involve commutators, which are zero for the $\eta'$.
\end{itemize}

The fundamental assumption of this paper is that the ``non-hairpin''
interactions involving pions and $\eta'$s are described by a Lagrangian
which has, at leading order in the chiral expansion,
the same form as the large $N_c$ Lagrangian, Eq. (\ref{chirallnc}).
The constants $f$ and $\mu$ will, however,
differ from those in both full QCD and the $N_c\to\infty$ limit.
There are additional ``hairpin'' interactions involving the $\eta'$,
and possibly pions, but not involving pions alone.
Hairpin interactions are those in which the quark and antiquark fields in
at least one of the external $\eta'$ mesons are contracted together.
Examples appear in Figs \ref{fthree}(d)-(i).
Before discussing these additional interactions,
I explain the justification for the assumption just announced.
This assumption has three parts.
\begin{enumerate}
%%%%%%%%%%%%%%%%%%%%%%%%%%%%%%%%%%%%%%%%%%%%%%%%%%%%%%%%%%%%%%%%%%%%%%%%%%%
\item
Interactions involving different numbers of non-singlet pions,
but no flavor singlet mesons,
are related to one another in the same way as in full QCD.
For example, the pion scattering lengths in the chiral limit
can be expressed in terms of $f_\pi$
according to the formulae of Weinberg \cite{weinberg}.
In full QCD the form of the chiral Lagrangian follows from three inputs
\cite{physica}:
(a) Ward identities between correlation functions;
(b) pion domination of low momentum amplitudes; and
(c) analyticity and unitarity of the amplitudes.
In the quenched approximation one has only the first two ingredients:
the Ward Identities in the quenched approximation are the same
as in full theories,
at least in lattice regularized theories \cite{wilsonWI,toolkit};
and pion dominance is a good approximation in practice.
But the amplitudes are neither analytic nor unitary.
In fact, it is only analyticity that is important
for the $O(p^2)$ terms in the Lagrangian;
the amplitudes from these terms are not themselves unitary.
What is required is that multi-pion amplitudes have a
polynomial dependence on the kinematic invariants.
This is the real assumption being made concerning the quenched approximation.
For staggered fermions, it is supported by numerical evidence
for amplitudes involving up to four pions \cite{pipinew}.
This is actually sufficient for the chiral logs discussed here,
which require only the two and
four point vertices from the $O(p^2)$ terms in the Lagrangian.
%%%%%%%%%%%%%%%%%%%%%%%%%%%%%%%%%%%%%%%%%%%%%%%%%%%%%%%%%%%%%%%%%%%%%%%%%%%
\item
Non-hairpin interactions of $\eta'$ mesons with pions
come entirely from $\cl_m$,
and are related to interactions of pions with each other by flavor factors.
In the large $N_c$ limit this result follows because the quark diagrams
are the same with or without $\eta'$ mesons.
This equivalence of diagrams is also true in the quenched approximation.
In particular, there are no internal quark loops in either the large $N_c$
limit or the quenched approximation.
Interactions involving intermediate gluons, which vanish when $N_c\to\infty$,
involve hairpins, and are discussed below.
This distinction is useful because it turns out that,
for some quantities, such hairpin interactions do not contribute
at leading order. This will become clearer in the following sections.
Thus I assume that the form of the effective Lagrangian describing the
pion-$\eta'$ interactions is the same in the quenched and $N_c\to\infty$
theories. The coefficients in the Lagrangians will differ, however,
because the gluon interactions are different.
%%%%%%%%%%%%%%%%%%%%%%%%%%%%%%%%%%%%%%%%%%%%%%%%%%%%%%%%%%%%%%%%%%%%%%%%%%%
\item
The interaction of the $\eta'$ with itself,
through both the the kinetic and mass terms,
is related to the self-interactions of pions by flavor factors.
This applies only to non-hairpin interactions resulting from diagrams
in which the $\eta'$ does not annihilate into gluons.
The justification for this assumption is as for the previous one:
the result is true in the large $N_c$ limit,
and the quenched diagrams involving $\eta'$ mesons are related to
those involving pions by the same flavor factors as in the large $N_c$ limit.
In particular, the mass eigenstates are the same as in that limit.
%%%%%%%%%%%%%%%%%%%%%%%%%%%%%%%%%%%%%%%%%%%%%%%%%%%%%%%%%%%%%%%%%%%%%%%%%%%
\end{enumerate}
An important consequence of the assumption that the
quenched Lagrangian has the same form as that for $\cl(N_c\to\infty)$
is that the non-singlet chiral currents are also the same.
As explained above,
this means that the $\eta'$ does not couple to these currents.

Although the form of the quenched Lagrangian is similar to that for full QCD,
the calculation of loop corrections using chiral perturbation theory
involves additional rules.
According to these rules certain loop diagrams are kept,
while others are discarded.
The chiral perturbation theory diagrams which are kept are those
corresponding to quark diagrams which do not contain internal quark loops.
The method by which one determines the correspondence between pion
and quark diagrams is best explained by example,
and is the subject of the following Section.

I now return to the additional hairpin interactions involving the $\eta'$.
It is here that the quenched and large $N_c$ theories differ:
such diagrams exist for the quenched approximation,
but vanish in the $N_c\to\infty$ limit.
For example, there is a ``disconnected'' contribution to the $\eta'$
two-point function in which the quarks annihilate into gluons.
It is this diagram, and its iterates including any number of quark bubbles,
that give the $\eta'$ its additional mass in full QCD.
More generally, such diagrams give rise to the $\eta'$ dependence in the
potentials $V_0$, $V_1$, $V_2$ and $V_5$, and thus convert the
Lagrangian $\cl_0+\cl_m$ into one with the form of Eq. (\ref{chiralleta}).

In quenched calculations one must include the contribution from
$\eta'$ loops since the $\eta'$ is light.
This means that the additional vertices coming from the $V_i$ may be needed.
Thus the coefficients of chiral logs due to $\eta'$ loops are not,
in general, predicted in terms of the leading order coefficients.\footnote{%
%%%%%%%%%%%%%%%%%%%%%%%%%%%%%%%%%%%%%%%%%%%%%%%%%%%%%%%%%%%%%%%%%%%%%
In full QCD this problem does not arise because the $\eta'$ is heavy,
and does not give rise to chiral logs. The uncertainties only enter
through the $O(p^4)$ terms in the chiral Lagrangian.}
%%%%%%%%%%%%%%%%%%%%%%%%%%%%%%%%%%%%%%%%%%%%%%%%%%%%%%%%%%%%%%%%%%%%%
They depend on new parameters which even in full QCD are either poorly
constrained by phenomenology, or not known at all.
It turns out, however, that there is a class of quantities
for which the $\eta'$ does not enter at one-loop order,
so that the additional $\eta'$ interactions are not needed.
This class includes $f_\pi$ and the kaon B-parameters, in the limit
that all quarks are degenerate.
These ``safe'' quantities are the main focus of this article.

For the remaining quantities one must include $\eta'$ loops.
As I explain in Section \ref{seta'}, these loops do not obey
usual power counting rules of chiral perturbation theory,
according to which each loop brings an additional power of $m_\pi^2$.
Instead, the leading order chiral logarithms are not suppressed by $m_\pi^2$,
and come from diagrams with any number of $\eta'$ loops.
These diagrams are of a simple form, however,
and require only the additional $\eta'^2$ interactions.\footnote{%
%%%%%%%%%%%%%%%%%%%%%%%%%%%%%%%%%%%%%%%%%%%%%%%%%%%%%%%%%%%%%%%%%%%%
Strictly speaking there are also two terms involving $V_2$ which
contribute. (1) A mass term proportional to $V_2''(0) \eta'^2 \Tr(M)$.
Using arguments similar to those of the following Section,
one finds that the $\Tr(M)$ corresponds to an additional quark loop.
Since this loop is absent in the quenched approximation, this term should
not be included.
(2) A vertex proportional to $V_2'(0) \eta' \Tr(M\eta'\pi^2)$.
This vertex actually appears in Fig. \ref{fthree}(f), and does contribute
to mass renormalization. By redefining the $U$ field, however, one can
absorb $V_2'(0)$ into the constant $A$ \cite{bernardgolterman}.
Thus there is no loss of generality if one excludes this term.}
%%%%%%%%%%%%%%%%%%%%%%%%%%%%%%%%%%%%%%%%%%%%%%%%%%%%%%%%%%%%%%%%%%%
These can be parameterized as
\begin{equation}
\label{eta'twopoint}
\cl_{\eta'} = {A-1 \over 2} \del_\nu \eta' \del_\nu \eta' +
{m_0^2 A \over 2 } \eta'^2  \ ,
\end{equation}
where the first term comes $V_5$, the second from $V_0$.
The most important feature of $\cl_{\eta'}$ is that $m_0$ is a fixed scale,
of $O(\Lambda_{QCD})$, and does not vanish in the chiral limit.
The parameterization of Eq. (\ref{eta'twopoint}) is chosen so that,
if this vertex could be iterated as in full QCD, the $\eta'$ would get a mass
\begin{equation}
m_{\eta'}^2=m_0^2 + {2\mu m \over A} \ .
\end{equation}
In the approximation that $A=1$, the experimental $\eta$ and $\eta'$
masses imply that $m_0\approx 0.9$ GeV.
I use these values when making estimates of the
quantitative importance of $\eta'$ loops.

\section{CHIRAL LOGARITHMS IN $f_\pi$ AND $f_K$}
\label{sfpi}
%%% sec3.tex
%%% section 3
%% \section{CHIRAL LOGARITHMS IN $f_\pi$ AND $f_K$}
%% \label{sfpi}

I begin by summarizing the calculation
of the chiral logarithmic corrections to the decay constant
in a theory with N light flavors, with quark loops included.
In chiral perturbation theory there are two one-loop diagrams,
and they are shown, along with the leading order diagram, in Fig. 1.
The circle represents the four-pion vertex obtained by expanding
Eq. (\ref{chirall}).
The square represents the axial current $\half(L_{\mu a}\!-\!R_{\mu a})$,
the expansion of which is as in Eq. (\ref{leftcurrent}),
except that only terms with an odd number of pions appear.
The crucial point about these vertices is that they are completely
determined in terms of the lowest order parameters $f$ and $\mu$.

%%% Fig 1
\begin{figure}
\vspace{1.5truein}
%\special{psfile=fig1.ps hoffset=24 voffset=12 hscale=0.9 vscale=0.9}
% for uw
\includegraphics{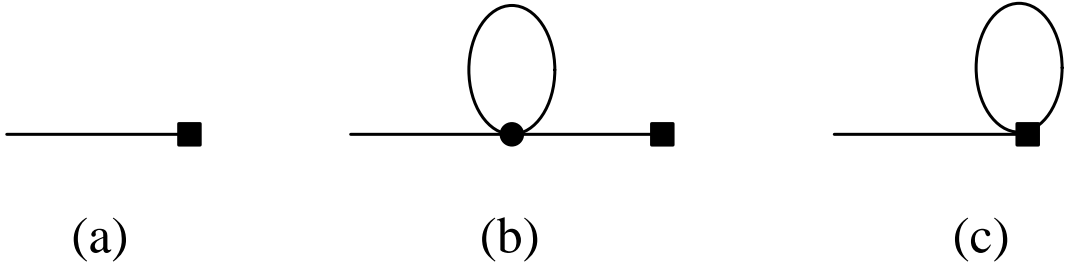}
\caption[fone]{\fc
Diagrams contributing to $f_\pi$: (a) leading order;
(b) wavefunction renormalization; (c) vertex renormalization.
The lines represent pseudo-Goldstone bosons, the box is the axial current,
and the circle is the vertex coming from $\cl$. }
\label{fone}
\end{figure}

As explained in the previous section, the leading order result is
$f({\pi_d})=f$, for all flavors $d$.
The corrections from Figs. \ref{fone}(b) and \ref{fone}(c) are multiplicative
\begin{equation}
\label{fpilogeqn}
f({\pi_d}) = f (1 + X({\rm Fig. \ref{fone}(b)})
                 + X({\rm Fig. \ref{fone}(c)}) +O(m_\pi^2) ) \ .
\end{equation}
The non-logarithmic corrections proportional to $m_\pi^2$ come from
$O(p^4)$ terms in the chiral Lagrangian.
They do not give rise to finite volume dependence (see Section \ref{sfinite}),
and I will drop them from subsequent equations.

Figure \ref{fone}(b) gives rise to both mass and wavefunction
renormalization, but only the latter effects $f_\pi$.
To contribute to wavefunction renormalization,
the vertex in Fig. \ref{fone}(b) must contain derivatives,
and thus must come from $\cl_0$.
This is an important observation to which I return below.
The contribution of Fig. \ref{fone}(b) is
\begin{equation}
\label{Xdef}
X({\rm Fig. \ref{fone}(b)}) =  {1\over 3 f^2}
\sum_c \Tr([T_c,[T_c,T_d]]T_d) \int_k G(k,m_c) \ ,
\end{equation}
where the measure is  $\int_k=\int d^4k /(2\pi)^4$,
and the integrand is the pion propagator $G(k,m)=1/(k^2+m^2)$.
The sum runs over the $N^2\!-\!1$ pions, and is to be done in the
mass eigenstate basis, in which the mass of the pion of flavor $T_c$
is labeled $m_c$.

Figure \ref{fone}(c) renormalizes the vertex;
its contribution turns out to
be proportional to wavefunction renormalization
\begin{equation}
X({\rm Fig. \ref{fone}(c)}) = - 4 X({\rm Fig. \ref{fone}(b)}) \ .
\end{equation}
Thus the fractional correction to the decay constant is
\begin{equation}
{f({\pi_d}) - f \over f} =  -
\sum_c \Tr([T_c,[T_c,T_d]]T_d)\ I_1(m_c) \ .
\end{equation}
Here I have defined one of the standard integrals that appear at one-loop
\begin{equation}
I_1(m) = {1\over f^2} \int_k G(k,m) \ .
\end{equation}
The integral must be regulated, and I use a sharp cut-off, $|k|<\Lambda$.%
\footnote{%
%%%%%%%%%%%%%%%%%%%%%%%%%%%%%%%%%%%%%%%%%%%%%%%%%%%%%%%%%%%%%%%
As is well known, such a momentum cut-off violates chiral symmetry.
This can be corrected by adding a quartically divergent counterterm,
but this does not contribute to any of the processes considered here
\cite{redfield}.}
%%%%%%%%%%%%%%%%%%%%%%%%%%%%%%%%%%%%%%%%%%%%%%%%%%%%%%%%%%%%%%%
The result is
\begin{equation}
I_1(m) = {\Lambda^2 \over (4\pi f)^2} + L(m) +
O\left({m^2\over\Lambda^2}\right) \ ,
%L(m) = \left({m \over 4\pi f}\right)^2 \ln\left({m^2\over\Lambda^2}\right)
\end{equation}
where $L(m)$ is chiral logarithm defined in Eq. (\ref{chirallogdef}).
The quadratic divergence can be absorbed by a redefinition of $f$.

For QCD, the general formula gives
\begin{eqnarray}
f_K &=& f\big(1 - \frac38 [2 I_1(m_K) + I_1(m_\eta) + I_1(m_\pi)] \big) \\
%    &&\stackrel{\longrightarrow}{m_u=m_d=0} f (1 - \frac54 L(m_K)) \ ; \\
f_\pi &=& f\big(1 - \frac12 [I_1(m_K) + 2 I_1(m_\pi)] \big) \ ,
%    &&\stackrel{\longrightarrow}{m_u=m_d=0} f (1 - \frac12 L(m_K)) \ ,
\end{eqnarray}
in agreement with the Refs. \cite{langackerpagels,gassleut,bbgfpi}.
For $N$ degenerate quarks the result is
\begin{equation}
\label{fpiNflavors}
f_\pi = f(1-\frac{N}2 I_1(m_\pi)) \ ,
\end{equation}
in agreement with Ref. \cite{condfinitev}.
%It is significant that the correction is proportional to $N$,
%as I discuss below.

I now turn to the quenched theory.
The result of the following discussion is extremely simple:
except for $\eta'$ loops,
which only contribute if the quarks are not degenerate,
there are {\em no leading order chiral logs} in decay constants.
I explain this conclusion in gory detail, as the insight obtained
can be applied directly to more complicated calculations.

The chiral logarithm comes from the infrared part of the loop integral,
$|p|< m$. The coefficient of the logarithm is determined by the residue
of the pole at $p^2=-m^2$.
In general this residue is itself determined by a physical,
on-shell amplitude: for wavefunction renormalization it is the
part of the four-pion scattering amplitude proportional to
the kinematic variable $t$, evaluated at threshold;
for vertex renormalization it is part of the amplitude for the axial current
to convert one pion to two pions.
In other words, to calculate the coefficient of the logarithm one can
factorize the loop in Figs. \ref{fone}(b) and (c) into two pieces:
an on-shell pion propagator,
and a physical scattering or conversion amplitude.
Since these amplitudes are evaluated for small $p$,
they are well described by lowest order chiral perturbation theory,
at least for small enough $m$.
This is why the coefficient of the log can be calculated in terms
of only the coefficients of the lowest order Lagrangian.

It is because of this factorization that one can hope to calculate the
coefficient of the logarithm in the quenched approximation.
As discussed in the previous section, it is reasonable to assume that
the leading order expressions for both the scattering amplitudes
and the chiral currents are the same in the quenched approximation as in
the full theory, except for differences in the coefficients $f$ and $\mu$.
Thus the coefficient of the chiral logs coming from each diagram
is the same function of these coefficients.
In other words, the calculation of individual chiral perturbation theory
diagrams proceeds by the same method in both theories.
The only differences in the quenched calculation are that
the $\eta'$ can propagate in the loops, and
those diagrams which involve internal quark loops should be discarded.

\begin{figure}
\vspace{1.5truein}
%\special{psfile=fig2.ps hoffset=144 voffset=12 hscale=0.7 vscale=0.7}
% for uw
\includegraphics{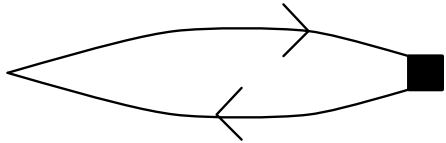}
%%\makebox[\textwidth][c]{(a) \hskip 3truein (b)}
\caption[ftwo]{\fc
Quark diagram corresponding to Fig. 1(a).
The lines represent quark propagators, as described in the text.
The box is the axial current. The apex where the propagators meet
is the operator which is used to create the pion.
}
\label{ftwo}
\end{figure}

To determine what these differences mean in practice it is necessary to find
the quark diagrams which correspond to the diagrams of chiral perturbation
theory.
Figure \ref{ftwo} shows the quark diagram corresponding to Fig. \ref{fone}(a).
The quark lines are to be interpreted as follows.
If one were doing a lattice (or any other non-perturbative) calculation of
$f_\pi$ one would calculate the diagram of Fig. \ref{ftwo},
with the lines corresponding to quark propagators in a background gauge field.
The result would then be averaged over the all background fields,
using either the full measure including quark loops, or the quenched measure.
The resulting value of $f_\pi$ would be correct to all orders
in $m_\pi^2$, and in particular would include the chiral logarithms.
In fact, the lines in Fig. \ref{ftwo} are almost the full propagators,
except that the propagation is
restricted so that there is no long distance pion contribution.
If the full measure is used, this restriction applies
also to the quarks loops contained in the background field.
With this restriction, the diagram does not produce chiral logarithms.
It does contribute terms of all orders in $m_\pi^2$,
but we are interested only in the lowest order contribution.

%%% Fig 2

This provides a definition of the meaning of the lines in quark diagrams
that is both vague and difficult to implement.
These shortcomings are not important, however, because the diagrams serve only
as a method for tracing the flow of the flavors of the quark and antiquark.
This aspect of the diagrams is completely well defined.

\begin{figure}
\vspace{4.0truein}
%\special{psfile=fig3.ps hoffset=12 voffset=8 hscale=0.8 vscale=0.8}
% fow uw
\includegraphics{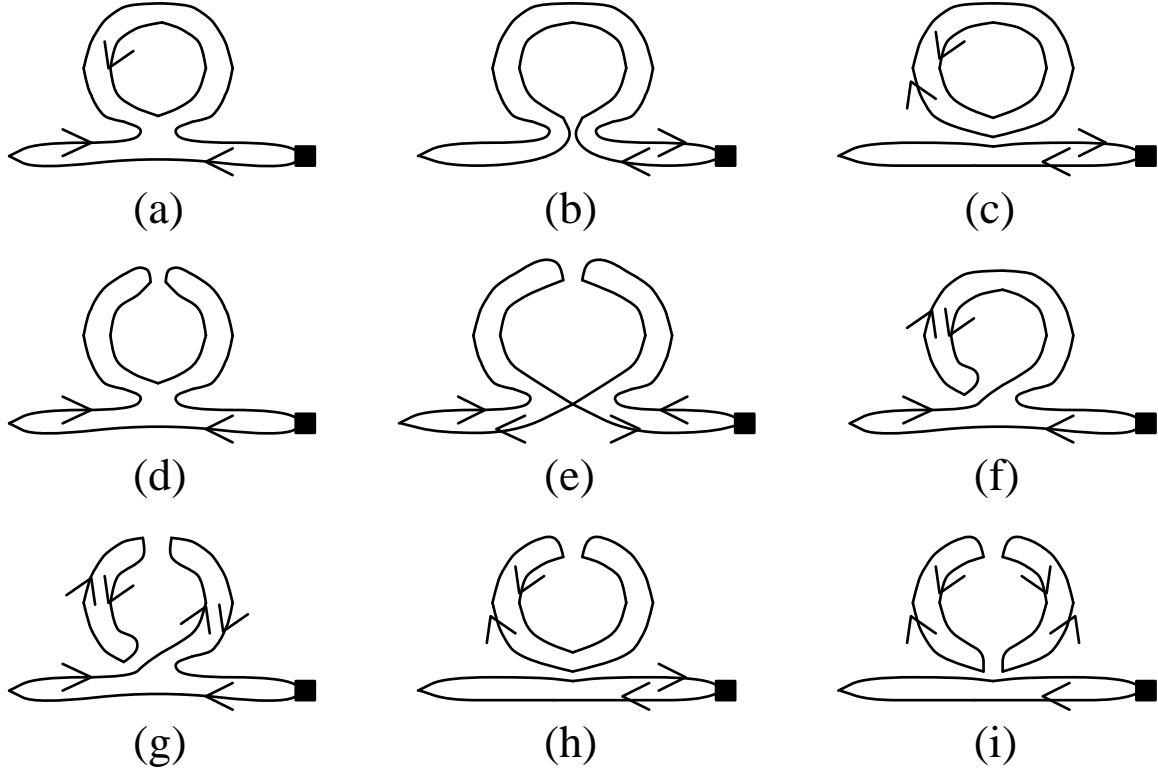}
%%\makebox[\textwidth][c]{(a) \hskip 3truein (b)}
\caption[fthree]{\fc
Quark diagrams corresponding to Fig. 1(b).
}
\label{fthree}
\end{figure}

The quark diagrams contributing to wavefunction renormalization
(i.e Fig. \ref{fone}(b)) are shown in Fig. \ref{fthree}.
These diagrams are obtained as follows.
First, draw all possible quark diagrams for the pion scattering amplitude
in which at least two of the pions have the flavor under consideration.
One of these two is the external pion, the other couples to the axial current.
Next, join the remaining two pions from the scattering amplitude
with a pion, or an $\eta'$, propagator.
Figure \ref{fthree} shows the topologies of the diagrams that result.
They represent the different ways in which flavor can flow in the
two step process of pion scattering followed by pion propagation.
It is only because the coefficient of the chiral log is determined
by this factorized limit that the flavor flow is well defined.

%%% Fig 3

Having understood the meaning of the diagrams in Fig. \ref{fthree},
I now analyze them in turn. The results of this analysis are
summarized in Table \ref{twflogs}.
The conclusion is that, in the full theory, only Fig. \ref{fthree}(a)
gives leading order chiral logs in wavefunction renormalization.
This means that it alone must correspond to the chiral
perturbation theory diagram, Fig. \ref{fone}(b).
In the quenched approximation, however, this diagram is absent.
Furthermore, if the quarks are degenerate,
none of the other diagrams contribute leading order quenched chiral logs.
Thus wavefunction renormalization does not contain leading order chiral logs
in the quenched approximation.

\begin{table}
% tabbing stuff
\begin{center}
\begin{tabular}{ccc}
Diagram & Full Theory ($N\ge4$)	&  Quenched Approximation \\ \hline
(a)	& Leading	& Absent (no loop)	\\
(b)	& Non-leading	& Non-leading		\\
(c)	& Non-leading	& Absent (no loop)	\\
(d),(e)	& Absent (no $\eta'$)& Proportional to $\delta m_q$ \\
(f)	& Absent (no $\eta'$)& Non-leading	\\
(g),(h),(i)& Absent (no $\eta'$)& Absent (no loop) \\
\end{tabular}
\end{center}
\caption[twflogs]
{
Contributions of the diagrams of Fig. \ref{fthree}
to chiral logarithms in wavefunction renormalization.
``Leading'' indicates a term proportional to $m_\pi^2\ln(m_\pi)$,
``non-leading'' a term proportional to $m_\pi^4\ln(m_\pi)$.}
\label{twflogs}
\end{table}

I begin by analyzing the diagrams in full QCD.
It turns out that
Figs. \ref{fthree}(d)-(i) do not contribute leading order chiral logs.
This is because they all contain one or more mesons annihilating into gluons.
This can only occur if the meson is the $\eta'$,
because at leading order the gluonic amplitude is flavor symmetric.
But in full QCD the $\eta'$ is massive and its loops do not give chiral logs.
The only possible contributions are thus from Figs. \ref{fthree}(a)-(c).
As I now explain, however, Figs. \ref{fthree}(b) and (c)
give rise only to non-leading chiral logs proportional to $m_\pi^4\ln(m_\pi)$.

\begin{figure}
\vspace{2.truein}
%\special{psfile=fig4.ps hoffset=24 voffset=12 hscale=0.9 vscale=0.9}
% for uw
\includegraphics{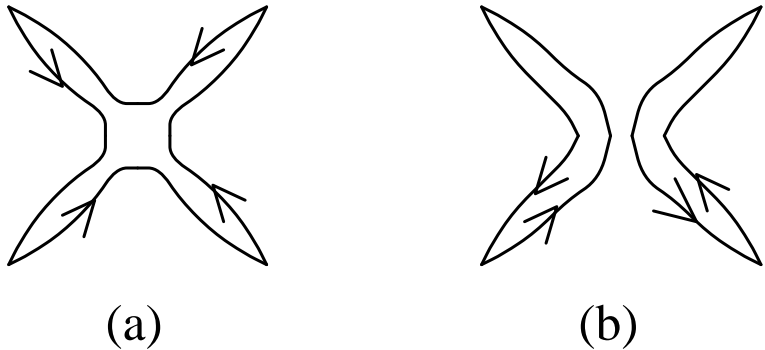}
%%\makebox[\textwidth][c]{(a) \hskip 3truein (b)}
\caption[ffour]{\fc
Quark diagrams contributing to pion scattering.
}
\label{ffour}
\end{figure}

The extra power of $m_\pi^2$ comes from the four-pion vertex.
The two types of diagram that contribute to pion scattering
are shown in Fig. \ref{ffour}:
diagram (a) is required for Fig. \ref{fthree}(a),
while diagram (b) is required for Figs. \ref{fthree}(b) and (c).
I claim that the amplitude from (a) is $O(p^2)$, while that from
(b) is $O(p^4)$.
(Here I am using $p^2$ loosely to refer to both $p^2$ and $m_\pi^2$.)
This is true both for the vertex coming from the derivative term in the
chiral Lagrangian (which is that required for wave function renormalization)
and for the vertex from the mass term.

%%% Fig 4

For a theory with four or more flavors this can be seen as follows.
One can choose the flavors of the external pions such that only
Fig. \ref{ffour}(b) contributes,
e.g. $\bar ud$, $\bar du$, $\bar sc$ and $\bar cs$.
For these flavors {\em the vertices from the leading order chiral Lagrangian
give no contribution}, because the flavor trace vanishes.
Only terms with two flavor traces contribute,
e.g. $\Tr(\del_\mu\Sigma \del_\mu \Sigma^{\dag})^2$,
but these contributions are of $O(p^4)$.
%In fact, the precise mass and momentum dependence is determined up to
%a few unknown coefficients.
For general external flavors, there will be contributions to the
scattering amplitude from both Figs. \ref{ffour}(a) and (b).
Those from the disconnected diagram (b) will, however, be of $O(p^4)$,
because the quark diagram is not affected if we simply change the flavor
labels.\footnote{%
%%%%%%%%%%%%%%%%%%%%%%%%%%%%%%%%%%%%%%%%%%%%%%%%%%%%%%%%%%%%%%%%%%%%%%%%%%
One might worry about the validity of this argument for non-degenerate quarks,
for then changing the flavor labels also means changing the quark masses.
This is not a problem, however, because there is always a theory in which,
for external flavors chosen so that only the disconnected diagram contributes,
the masses of the quarks are the same as those in the diagram under
consideration. In all such theories the disconnected components are $O(p^4)$.}
%%%%%%%%%%%%%%%%%%%%%%%%%%%%%%%%%%%%%%%%%%%%%%%%%%%%%%%%%%%%%%%%%%%%%%%%%%
Thus the only contribution of $O(p^2)$ can be from Fig. \ref{ffour}(a).
That this diagram does contribute at $O(p^2)$ can be seen by choosing
flavors such that it is the only contribution,
e.g. $\bar ud$, $\bar ds$, $\bar sc$ and $\bar cu$.
The $O(p^2)$ terms in the chiral Lagrangian do give a vertex with these
flavors, because they can be connected together in a single flavor trace.

This argument requires only that the leading order interactions
are described by the chiral Lagrangian.
Given the fundamental assumption of this paper, namely that the Lagrangian
for quenched interactions has the same form as in the full theory,
the argument of the previous paragraph applies also to the quenched theory.
In fact, it applies also if the external particles are flavor singlets.
Since the background configurations in the quenched approximation
are by construction independent of the number of flavors
of the full theory one starts with,
the argument also applies to quenched QCD.

An interesting question is whether Figs. \ref{fthree}(b) and (c) gives rise to
leading order chiral logs in full QCD. Since QCD has only three light flavors,
the argument given above does not work: none of the possible external
flavors selects Fig. \ref{ffour}(b) alone.
This means that the question cannot be asked in terms of correlation functions
of operators, but only in terms of individual contractions. Thus one must
have a non-perturbative regulator, such as the lattice.
Although I do not know of a proof,
I think it likely that the $N\ge4$ result carries over to $N=3$,
for the following reasons.
First, with staggered fermions, one can show,
given certain assumptions, that Fig. \ref{ffour}(b) is $O(p^4)$
irrespective of the number of dynamical quarks \cite{pipinew}.
Second, the contribution of wavefunction renormalization
in a theory with $N$ degenerate quarks, Eq. (\ref{Xdef}),
comes with an explicit factor of $N$ for $N\ge2$.
This suggests that Fig. \ref{fthree}(a) alone is contributing,
because there are $N$ flavors of quark running in the loop.
This is not conclusive, however, because there could be hidden
$N$ dependence in $f$ and $\mu$.
Finally, the claim that only Fig. \ref{fthree}(a) contributes
is consistent with the $N_c\to\infty$ limit,
in which the both the vertex from the leading order chiral Lagrangian
and that from Fig. \ref{fthree}(a) are of $O(1/N_c)$,
while those from Figs. \ref{fthree}(b) and (c) are of $O(1/N_c^2)$.
%%It would be interesting to test this claim in numerical simulations.

I now turn to the analysis of the quark diagrams
in the quenched approximation.
Of the first three diagrams, only Fig. \ref{fthree}(b) is present in the
quenched approximation, as the other two involve quark loops.
As I have just argued, however, Figs. \ref{fthree}(b) gives non-leading
chiral logs not only in the full theory (with $N\ge4$),
but also in the quenched approximation.
Thus these three diagrams do not contribute leading order chiral logs to
quenched $f_\pi$.

The remaining diagrams involve $\eta'$ propagation.
Of these, Figs. \ref{fthree}(g)-(i) require internal quark loops,
and so are not present in the quenched approximation.
Figures \ref{fthree}(d)-(f), however, do give potential contributions
of a type unrelated to those in full QCD.
Fortunately, if the quarks are degenerate, this contribution
vanishes at leading order.
This is because, as explained in Section \ref{schiral},
the $\eta'$ does not appear in the four-pion vertex containing derivatives.
(Recall that for wavefunction renormalization the vertex must contain
derivatives.)
For degenerate quarks, one can use the usual basis of $\pi_0$, $\eta$
and $\eta'$, and since the $\eta'$ does not couple,
Figs. \ref{fthree}(d)-(f) simply vanish.
There will, however, be contributions at higher order chiral in
perturbation theory.

In fact, Fig. \ref{fthree}(f) does not contribute leading order
chiral logs to $f_\pi$ even for non-degenerate quarks.
This is because only the $V_2$ term in the full chiral Lagrangian,
Eq. (\ref{chiralleta}), has the correct flavor
structure to correspond to the vertex in Fig. \ref{fthree}(f).
Since this term has no derivatives, however,
it cannot contribute to wavefunction renormalization
irrespective of the quark masses.

This is not true of Figs. \ref{fthree}(d) and (e).
For non-degenerate quarks, the neutral eigenstates have flavors
$\bar ss$, $\bar dd$ and $\bar uu$, and each of these individually
does couple to the derivative term in the Lagrangian.
It is simple to see that the contribution to
wavefunction renormalization vanishes for degenerate quarks
because of a cancellation between the two diagrams.
For non-degenerate quarks, this cancellation does not occur
because the neutral eigenstates are different.
Thus there are extra contributions proportional to the difference in mass
of the two quarks making up the pion.
For example, this extra contribution is proportional to $m_d-m_u$
for the physical $f_\pi$.
These extra terms are discussed in Ref. \cite{bernardgolterman},
and I do not consider them further here.

For mass renormalization Figs. \ref{fthree}(d-f) all contribute
even if the quarks are degenerate,
as discussed further in Section \ref{seta'}.

\begin{figure}
\vspace{3.0truein}
%\special{psfile=fig5.ps hoffset=48 voffset=8 hscale=0.9 vscale=0.9}
%% for uw
\includegraphics{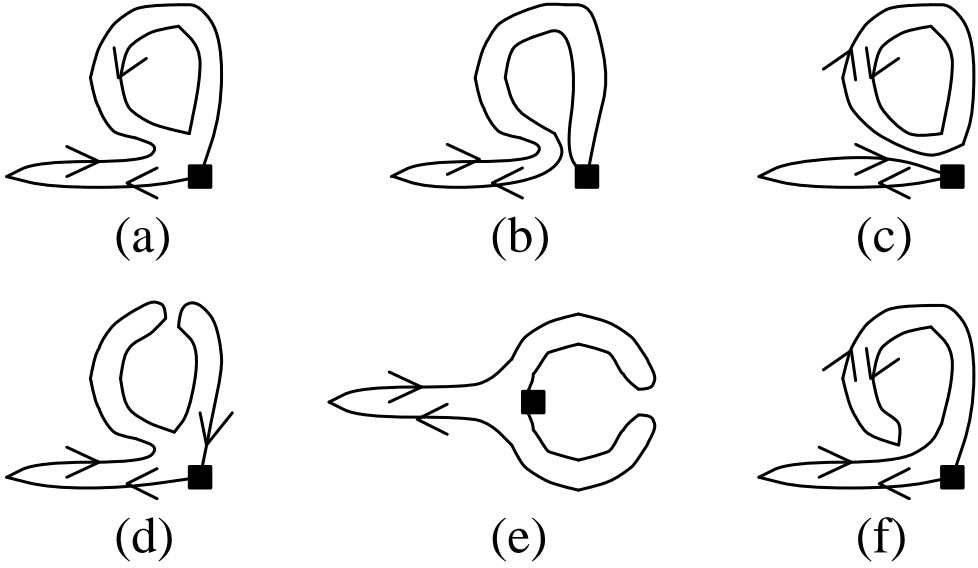}
%%\makebox[\textwidth][c]{(a) \hskip 3truein (b)}
\caption[ffive]{\fc
Quark diagrams corresponding to Fig. 1(c).
}
\label{ffive}
\end{figure}

Finally, I turn to vertex renormalization, Fig. \ref{fone}(c).
The quark diagrams are in one-to-one correspondence with
those in Fig. \ref{fthree}, and the analysis is almost identical.
The result is that the summary of Table \ref{twflogs} applies
both to vertex and wavefunction renormalization.
In Fig. \ref{ffive} I show the diagrams analogous to
Figs. \ref{fthree}(a)-(f).
I do not show those corresponding to Figs. \ref{fthree}(g)-(i),
since they contribute neither to full nor to quenched theories.
Fig. \ref{ffive}(a) is the only diagram contributing leading order chiral
logs in the full theory, but is absent in the quenched approximation.
The three pion vertex in Figs. \ref{ffive}(b) and (c)
is absent in leading order because it requires two flavor traces,
whereas the axial current contains a single flavor trace
(Eq.  (\ref{leftcurrent})).
Thus these diagrams contribute only non-leading chiral logs.
This is true also in the quenched theory,
for which only diagram (b) is present.
The remaining three diagrams require an $\eta'$ in the loop,
and are thus absent in the full theory.
They are present in general in the quenched theory, but vanish for
degenerate quark masses because, as explained in Section \ref{schiral},
the $\eta'$ does not couple to the leading order representation of
the axial current.
% Eq. (\ref{leftcurrent}).

%%% Fig 5

In summary, the leading order chiral logs in decay constants,
which are present in full QCD, are absent in the quenched approximation.
There are additional chiral logs from $\eta'$ loops,
which depend on parameters other than $f$ and $\mu$,
but these are only present for non-degenerate quarks.
These extra logs are discussed in Ref. \cite{bernardgolterman}.

\section{CHIRAL LOGARITHMS IN B-PARAMETERS}
\label{sbk}
%%% sec4.tex
%%% section 4
%% \section{CHIRAL LOGARITHMS IN B-PARAMETERS}
%% \label{sbk}
\subsection{$B_K$ with dynamical quarks}

To set the stage for the calculation of the chiral logs in $B_K$
in the quenched approximation, I review the calculation in theories
with dynamical quarks. The discussion will apply to theories with
a $d$ quark, an $s$ quark, and any number of other light quarks.

The kaon $B$-parameter is defined as a ratio of matrix elements
\begin{eqnarray}
\label{firstbk}
\cm_K &=& \vev{\bar{K^0}|\co_K|K^0} = B_K \cm_\vac \ , \\
\co_K &=& [\bar{s}_a\g\mu\pl d_a][(\bar{s}_b\g\mu\pl d_b)] \ ,\\
\cm_\vac &=& \frac83 \vev{\bar{K^0}|[\bar{s}_a\g\mu\pl d_a]|0}
                      \vev{0|[(\bar{s}_b\g\mu\pl d_b)]|K^0} \ ,
\end{eqnarray}
where $a$ and $b$ are color indices.
The subscript ``vac'' indicates that this is the result for $\cm_K$
in vacuum saturation approximation.
%%% In the vacuum saturation approximation $B_K=1$.
The definition can be written in the more familiar form
\begin{equation}
\label{secondbk}
\cm_K = \frac{16}{3} B_K f_K^2 m_K^2 \ .
\end{equation}
The calculation of chiral logs is clearer using the definition
in Eq. (\ref{firstbk}).
%%% The momenta are both incoming,
%%% so that $p_{\bar K}\cdot p_K= - p_K^2 = m_K^2$.
What is required is the logs in both $\cm_K$ and $\cm_\vac$.

I begin with the latter.
The matrix element $\cm_\vac$ is the product of two matrix elements of the
axial current, the corrections to which are discussed in the previous Section.
The result can be read off from Eqs. (\ref{fpidef}) and (\ref{fpilogeqn})
\begin{equation}
\label{vaclogs}
\cm_\vac = \frac{16}{3} f^2 p_{\bar K}\cdot p_K
(1 + 2 X({\rm Fig. 1b}) + 2 X({\rm Fig. 1c}) + O(m_K^2) ) \ .
\end{equation}
The $O(m_K^2)$ term comes from non-leading operators, and does not
contain chiral logs.
As in the previous Section, I will drop such terms henceforth.
In the formula for $X$ (Eq. \ref{Xdef}) one should use the flavor matrix
appropriate to the kaon, $T_d=\frac{1}{\sqrt2}(T_6\!-\!iT_7)$.
The momenta are both incoming,
so that $p_{\bar K}\cdot p_K= - p_K^2 = m_K^2$,
where $m_K$ is the physical mass including one-loop corrections.
%I prefer, however, to keep the momenta explicit.
%This avoids the suggestion that the loop corrections to $m_K^2$
%need to be taken into account, which they do not.
The corrections in Eq. (\ref{vaclogs}) are precisely those
required to change $f^2$ to the one-loop corrected $f_K^2$
\begin{equation}
\label{mvacres}
\cm_\vac = \frac{16}{3} f_K^2 m_K^2 \ .
\end{equation}

The chiral logarithmic corrections to $\cm_K$ in full QCD were first
calculated in Ref. \cite{sonoda},
and related results are given in Ref. \cite{buras}.
I have checked these calculations, and in the following I present
a slight generalization of the results.
The $\Delta S=2$ operator $\co_K$ transforms as part of a
$(27,1)$ under $SU(3)_L\times SU(3)_R$.
In chiral perturbation theory it is represented by a sum of all operators
having the same transformation property, with unknown coefficients.
There are no such operators at $O(1)$,
and only one at $O(p^2,m_q)$ \cite{sonoda}
\begin{equation}
\label{cokchieqn}
\co_K \to \co_K^{\chi} =
\frac{4}{3} B f^4 (\Sigma\del_\mu\Sigma^{\dag})_{ds}
                  (\Sigma\del_\mu\Sigma^{\dag})_{ds} \ ,
\end{equation}
where the subscripts indicate the $ds$'th element of the matrix.
I have chosen the coefficients so that, at tree level, $B_K=B$,
as can be seen by combining the tree level result,
\begin{equation}
\cm_K = \frac{16}{3} B f^2 m_K^2 \ ,
\end{equation}
with that for $\cm_\vac$ from Eq. (\ref{vaclogs}).
Chiral perturbation theory does not determine the value of $B$,
which is another parameter like $f$ and $\mu$.

\begin{figure}
\vspace{1.5truein}
%\special{psfile=fig6.ps hoffset=12 voffset=-236 hscale=0.8 vscale=0.8}
% for uw
\includegraphics{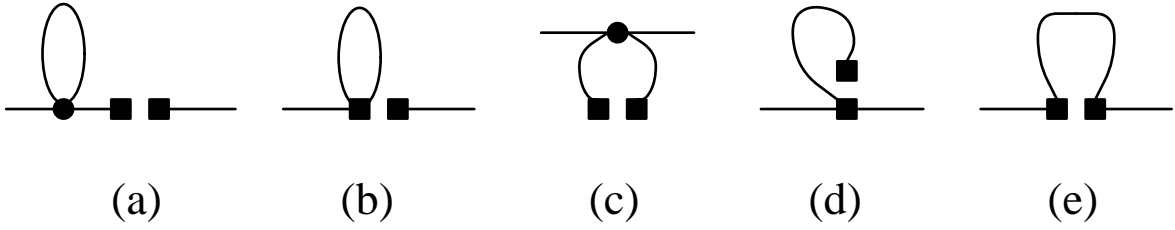}
%%\makebox[\textwidth][c]{(a) \hskip 3truein (b)}
\caption[fsix]{\fc
Chiral perturbation theory diagrams for $\cm_K$. The pair of squares
represents the operator $\co_K^{\chi}$, while the circle is the vertex
from the chiral Lagrangian.
}
\label{fsix}
\end{figure}

Chiral logs in $\cm_K$ come from the five types of diagrams
shown in Fig. \ref{fsix}.
I have represented the operator by two squares,
each corresponding to a left-handed current, because it can be written as
\begin{equation}
\co_K^{\chi} =
\frac{16}{3} B (L_{\mu 6}\!-\!i L_{\mu 7})(L_{\mu 6}\!-\!i L_{\mu 7}) \ .
\end{equation}
This factorization, which plays an important role in the calculation,
is only true at leading order in the chiral expansion.
The result from these diagrams is a multiplicative correction
\begin{equation}
\label{fulllogs}
\cm_K = \frac{16}{3} B f^2 m_K^2
\left(1+X({\rm Fig.6(a)}) +
X({\rm Fig.6(b)}) + X({\rm Fig.6(c\!-\!e)})\right) \ .
\end{equation}
Here $m_K$ is the one loop corrected kaon mass.
Strictly speaking, some of the results from Figs. \ref{fsix}(c)-(e) are
proportional to the tree level value of $m_K^2$
(or other tree-level meson masses squared).
But since these are correction terms, one is justified in replacing
$m_K({\rm tree})$ with $m_K({\rm 1 loop})$.

%%% Fig 6

The calculation of Figs. \ref{fsix}(a) and (b) is identical to that of
Figs. \ref{fone} (b) and (c) respectively,
except for an overall factor of two since there are two left handed currents
in $\co_K^\chi$
\begin{equation}
\label{Xa-b}
X({\rm Fig.6(a)})= 2 X({\rm Fig.1(b)})\, ;\ \
X({\rm Fig.6(b)})= 2 X({\rm Fig.1(c)})\ .
\end{equation}
This part of the correction to $\cm_K$ is the same as the complete
correction to $\cm_\vac$, Eq. (\ref{vaclogs}),
and so cancels in the ratio which defines $B_K$.
In other words, the corrections from Figs. \ref{fsix}(a) and (b) are
precisely those needed to convert the factor of $f^2$ in Eq. (\ref{fulllogs})
to the one loop value $f_K^2$, and thus do not affect $B_K$.
This is true for any number of flavors in addition to the $s$ and $d$
quarks.

The only new work is that required to calculate Figs. \ref{fsix}(c)-(e).
The result can be written in a form which is valid for any number of flavors
\begin{equation}
\label{Xc-e}
X({\rm Fig.6(c\!-\!e)}) = I_2(m_K) -
\sum_a (1+{m_a^2\over m_K^2}) I_1(m_a)
\left[(T_a)_{dd} - (T_a)_{ss}\right]^2 \ ,
\end{equation}
where $I_2$ is the standard integral
\begin{eqnarray}
I_2(m) &=& {m^2\over f^2} \int_k G(k,m)^2 \\
       &=& - L(m) - \left({m\over 4\pi f}\right)^2(1 + O(m^2/\Lambda^2)) \ .
\end{eqnarray}
This integral results from Figs. \ref{fsix}(c) and (d),
both of which have a $K^0$ in the loop.
In Fig. \ref{fsix}(e) there are only flavor-diagonal pions in the loop.
This diagram gives rise to the second term in $X$,
in which the sum is over flavor-diagonal pions
in the mass eigenstate basis.\footnote{%
%%%%%%%%%%%%%%%%%%%%%%%%%%%%%%%%%%%%%%%%%%%%%%%%%%%%%%%%%%%%%%%%%%%%%%%
In fact, the sum can be extended to all pions,
because the last factor is non-zero only for flavor-diagonal pions.
The same is true in Eq. (\ref{correctedBKeqn}) below.}
%%%%%%%%%%%%%%%%%%%%%%%%%%%%%%%%%%%%%%%%%%%%%%%%%%%%%%%%%%%%%%%%%%%%%%%

An important consistency check on this result is that divergent parts
can be absorbed into $B$.
This requires that there be no quartic divergences, for these would
imply that $\cm_K \propto \Lambda^4$, while chiral symmetry dictates
that the leading term is proportional to $m_K^2$.
All three diagrams do contains quartic divergences,
but they cancel in the sum.
There are quadratic divergences in Eq. (\ref{Xc-e}) because
$I_1 \propto \Lambda^2$ at leading order.
These can be absorbed into $B$ as long as they
do not depend on the quark masses.
The factor of $m_a^2/m_K^2$ multiplying $I_1(m_a)$ suggests that this
may not be true, but in fact the mass dependence cancels because of the result
\begin{equation}
\sum_a m_a^2 \left[(T_a)_{dd} - (T_a)_{ss}\right]^2 =  m_K^2 \ .
\end{equation}
Here again the sum is in the mass eigenstate basis,
and only flavor-diagonal states contribute.
This result holds for any number of flavors.

Combining Eqs. (\ref{mvacres}), (\ref{fulllogs}), (\ref{Xa-b})
and (\ref{Xc-e}), the corrected $B_K$ is
\begin{equation}
\label{correctedBKeqn}
B_K = B \left(1 + I_2(m_K) - \sum_a (1+{m_a^2\over m_K^2}) I_1(m_a)
\left[(T_a)_{dd} - (T_a)_{ss}\right]^2 \right) \ .
\end{equation}
In QCD (with $m_u=m_d$) this gives
\begin{equation}
\label{bkfinalres}
B_K = B \left(1 + I_2(m_K) - \frac34 (1+{m_\eta^2\over m_K^2}) I_1(m_\eta)
         - \frac14 (1+{m_\pi^2\over m_K^2}) I_1(m_\pi) \right) \ .
\end{equation}
In the limit $m_u=m_d\to0$ the chiral logarithmic part is
\begin{equation}
B_K \longrightarrow B (1 - 4 L(m_K)) \ ,
\end{equation}
where the quadratic divergences have been absorbed into $B$.
It is also useful to expand about $m_s=m_d$,
in terms of $\delta=(m_s\!-\!m_d)^2/(m_s\!+\!m_d)^2$
\begin{equation}
B_K =  B (1 - L(m_K)(3 + \frac13 \delta) + O(\delta^2)) \ ,
\end{equation}
a result previously quoted in Ref. \cite{book}.
Finally, for $N$ degenerate flavors one finds
\begin{equation}
\label{bkdegenerate}
B_K =  B (1 + I_2(m) -2 I_1(m)) = B (1- 3 L(m) + \dots ) \ .
\end{equation}
Unlike the correction to $f_\pi$,
this result has no explicit dependence on $N$.

\subsection{$B_K$ in the quenched approximation}

I now turn to the calculation of the chiral logs in quenched $B_K$.
The conclusion of the following analysis is simple:
the result for $B_K$ is unchanged from the full theory except that
(1) there are contributions from $\eta'$ loops when $m_s\ne m_d$, and
(2) in the formula Eq. (\ref{correctedBKeqn})
the mass eigenstates are $\bar ss$ and $\bar dd$
(with masses $m_{ss}= 2\mu m_s$ and $m_{dd}= 2\mu m_d$, respectively).
Thus the quenched result is
\begin{eqnarray}
\nonumber
B_K({\rm quenched}) &=& B \left(1 + I_2(m_K)
			- {3m_d+m_s \over 2m_d+2m_s} I_1(m_{dd})
			- {m_d+3m_s \over 2m_d+2m_s} I_1(m_{ss}) \right. \\
\label{bkquenched}
&&\qquad\qquad
			\left. + (m_s-m_d)X(\eta') \right) \ ,
\end{eqnarray}
where $X(\eta')$ comes from $\eta'$ loops.
In the limit $m_d=m_s$ the result is identical to that in the full theory,
Eq. (\ref{bkdegenerate}).

To begin the quenched calculation,
the operator $\co_K$ must be written in terms of pion fields.
As before, I assume that the form of the operator is the same as
for the full theory, Eq. (\ref{cokchieqn}), except that the constant $B$
can differ. In other words, I assume that the amplitudes for the operator to
create two and four pions are related in the same way as in the full theory.
In the continuum, these relations are derived from the chiral Ward identities.
With staggered fermions, analogous Ward identities hold in both full and
quenched theories \cite{toolkit,book}.
With Wilson fermions, the Ward identities are regained
in the continuum limit of both full and quenched theories \cite{wilsonWI}.
Dynamical fermions play no role in the derivations.

Given this assumption, the calculation proceeds as in the continuum,
except for the usual two complications.
First, one must keep only those diagrams in Fig. \ref{fsix} which
correspond to quark diagrams without internal loops.
Second, one must determine whether there are extra contributions from
$\eta'$ mesons propagating in the loops.

For Figs. \ref{fsix}(a) and (b) the analysis from the previous Section
can be used,
because the operator $\co_K^{\chi}$ is a product of two factors of $L_\mu$.
The calculation of Figs \ref{fsix}(a) and (b) is identical to that
of Figs. \ref{fone}(b) and (c).
This means that, just as in the full theory, any chiral logs from
Figs. \ref{fsix}(a) and (b) cancel in ratio which defines $B_K$.
In fact, as shown in the previous Section, the chiral logs in quenched
$f_K$ differ from those in the full theory.
In particular there are no logs in $f_K$ for $m_s=m_d$, and there
are extra logs from $\eta'$ loops if $m_s\ne m_d$.
But these differences are irrelevant for $B_K$.

\begin{figure}
\vspace{3.0truein}
%\special{psfile=fig7.ps hoffset=12 voffset=-132 hscale=0.8 vscale=0.8}
% for uw
\includegraphics{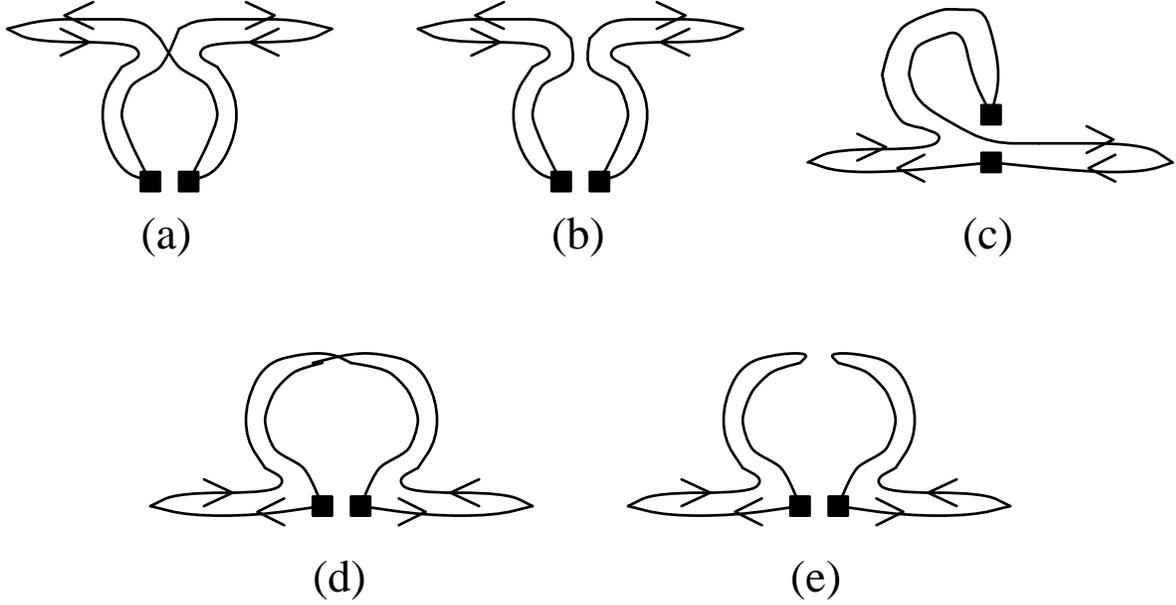}
%%\makebox[\textwidth][c]{(a) \hskip 3truein (b)}
\caption[fseven]{\fc
Quark diagrams corresponding to Figs. 6(c-e).
}
\label{fseven}
\end{figure}

Just as in the full theory, then,
the logs for $B_K$ come entirely from Figs. \ref{fsix}(c)-(e).
The quark diagrams corresponding to these are shown in Fig. \ref{fseven}.
The operator $\co_K$ is shown as two squares,
each representing a left handed current.
Each diagram actually represents two contractions, one in which
the color indices are contracted within each current, the other having
the indices contracted between currents.
The pairings of quark and antiquark propagators into bilinears
can be interchanged by a Fierz transformation.
I have chosen the pairing so that each bilinear couples only to one of the
mesons in the loop.
This makes it easier to see the correspondences with pion diagrams.

%%% Fig 7

None of the diagrams in Fig. \ref{fseven} contain internal quark loops.
Thus all are present in both quenched and full theories.
The correspondences between quark and pion diagrams is as follows.
Figures \ref{fseven}(a) and (b) correspond to Fig. \ref{fsix}(c),
since both have a four pion scattering amplitude.
Figure \ref{fseven}(c) corresponds to Fig. \ref{fsix}(d),
since the particle in the loop is a kaon.
And, finally, Figs. \ref{fseven}(d) and (e) correspond to Fig. \ref{fsix}(e),
in which the mesons in the loop have flavors $\bar ss$ and $\bar dd$.

Since Figs. \ref{fseven}(a), (b) and (c) are the same in both full
and quenched theories, and since the particles in the loops are flavor
off-diagonal, the result from the chiral perturbation theory diagrams
Figs. \ref{fsix}(c) and (d) is identical in full and quenched theories.
This is the first example in which there are chiral logs in a quenched
quantity coming from the same diagrams as those in the full theory.
In fact, Fig. \ref{fseven}(b) does not contribute at leading order
because the four pion vertex is $O(p^4)$.

The quenched approximation can, however, affect the evaluation of
Fig. \ref{fsix}(e), because the loop contains flavor-diagonal mass eigenstates,
which, in general, are different in quenched and full QCD.
There are also potential $\eta'$-loop contributions in the quenched theory
not present in the full theory, for example that of Fig. \ref{fseven}(e).
The effect of these differences is that, in the expression for
$X({\rm Fig.\ref{fsix}(c\!-\!e)})$, Eq. (\ref{Xc-e}),
the sum in the quenched theory runs over both the pions and the $\eta'$,
while in full QCD the $\eta'$ is excluded.
The difference between these two sums vanishes, however,
for degenerate quarks. This is for two reasons.
First, the non-singlet pion mass eigenstates are the same in quenched and
full theories, so that the contribution to the sums from these pions agree.
Second, the $\eta'$ is a mass eigenstate, and its extra contribution
to quenched sum vanishes, because $T_{ss} - T_{dd}=0$ for a flavor singlet.
Another way of seeing this is to recall that the $\eta'$ does not couple
to $L_\mu$, so the extra diagram \ref{fseven}(e) vanishes.

It follows from this discussion that Eq. (\ref{bkquenched}) describes the
leading order chiral logs in quenched $B_K$.
%It is important to realize that,
%apart from the $\eta'$ loops of Fig. \ref{fseven}(e),
%the quark diagrams in both full and quenched theories agree not only
%at leading order but also at higher orders.
%Thus some of the higher order logs will agree also.
%Only higher order logs coming from mass renormalization, which is affected
%by $\eta'$ loops, will disagree.

\subsection{$B_V$ and $B_A$}

The matrix element $\cm_K$ picks out the positive parity part of $\co_K$,
which consists of a sum of eight different terms:
four ``vector-vector'' ($\g\mu \cdot \g\mu$)
and four ``axial-axial'' ($\g\mu\g5 \cdot \g\mu\g5$).
There are two different color contractions
for each of these eight terms, with one or two traces over color indices.
When doing a numerical calculation one calculates all sixteen terms
separately, so it is interesting to predict their chiral behavior
and finite volume dependence.
In this subsection I calculate the leading order chiral logarithms for
the four Lorentz invariant combinations in which the index $\mu$ is summed.

The precise definition of the breakup of $\cm_K$ into sixteen pieces
is this:
\begin{enumerate}
\item Write down all contractions (i.e. products of quark propagators).
\item Fierz transform the operator such that both external kaons are
	connected to one or other bilinear. All terms will thus have
	two traces over spinor indices, and either one or two traces
	over color indices.
\item Break up each of the resulting terms into the eight
	vector-vector and axial-axial pieces.
\end{enumerate}
\noindent
Clearly, this definition applies at the level of contractions.
In order to use chiral perturbation theory, however,
one must have a definition in terms of matrix elements of operators.
For three or less flavors it is not possible to write down
operators whose matrix element give each of the sixteen terms separately.
As I show shortly, to do this requires at least four valence flavors.
In quenched QCD, this presents no problem,
for one has already decoupled the number of valence and dynamical flavors.
%The extra approximation is no different from that already made.
It does mean that the results derived below are not directly
applicable to full QCD.
They should, however, be as valid for full QCD as they are for quenched QCD.

Consider, then, a theory with $N\ge4$ light flavors:
an $s$ and an $s'$, both of mass $m_s$;
a $d$ and a $d'$, both of mass $m_d$;
and $N-4$ other quarks with arbitrary (though small) masses.
Let the $K$ be the $d\bar s$ pion, and the $K'$ be the corresponding
state made out of primed quarks.
Define $\cm_K$ and $B_K$ in this theory exactly as in QCD
\begin{eqnarray}
\cm_K 	&=&
\vev{\bar{K}|[\bar{s}_a\g\mu\pl d_a][(\bar{s}_b\g\mu\pl d_b)] |K^0} \ , \\
 	&=& \frac{16}{3} B_K f_K^2 m_K^2 \ .
\end{eqnarray}
Then, by comparing contractions, we can write $\cm_K$ as follows
\begin{eqnarray}
\label{ovdefeqn}
\cm_K     &=& 2(\cm_V + \cm_A) \\
\cm_V   \ &=& \cm_{V1} + \cm_{V2} \\
\cm_A   \ &=& \cm_{A1} + \cm_{A2} \\
\label{v1defeqn}
\cm_{V1} \ &=& \vev{\bar{K'}| [\bar s'_a \g\mu d'_b] \
                                [\bar s_b  \g\mu d_a]  |K} \\
\label{v2defeqn}
\cm_{V2} \ &=& \vev{\bar{K'}| [\bar s'_a \g\mu d'_a] \
                                [\bar s_b  \g\mu d_b]  |K} \\
\label{a1defeqn}
\cm_{A1} \ &=& \vev{\bar{K'}| [\bar s'_a \g\mu \g5 d'_b] \
                                [\bar s_b  \g\mu \g5 d_a]  |K} \\
\label{a2defeqn}
\cm_{A2} \ &=& \vev{\bar{K'}| [\bar s'_a \g\mu \g5 d'_a] \
                                [\bar s_b  \g\mu \g5 d_b]  |K}\ .
\end{eqnarray}
The point of the extra flavors is to restrict contractions;
each of the matrix elements in the last four lines has only one
contraction.
The factor of two on the first line appears because
the two bilinears in $\cm_K$ are identical,
and so there are twice as many contractions as in $\cm_{V}$ and $\cm_{A}$.
The subscripts $1$ and $2$ indicate the number of traces over color indices.

The corresponding $B$-parameters are defined by
\begin{equation}
\label{bvdefeqn}
\cm_{V1} = \frac83 f_K^2 m_K^2 B_{V1} \ ,\ \
\cm_{V2} = \frac83 f_K^2 m_K^2 B_{V2} \ ,
\end{equation}
and identical equations with $V\to A$.
These definitions are such that
\begin{equation}
B_K = B_V + B_A  = B_{V1} + B_{V2} + B_{A1} + B_{A2} \ .
\end{equation}
It is straightforward to show that in vacuum saturation approximation
$B_{A1}=0.25$, $B_{A2}=0.75$, and thus $B_A=1$,
while $B_V=B_{V1}=B_{V2}=0$.
The numerical results, shown in Section \ref{snumer},
do not, in fact, agree with these predictions.

I calculate the chiral logs in these $B$-parameters in the full theory,
and then argue that the results apply with only minor
changes in the quenched approximation.
The first step is to transcribe the problem into the language of
chiral perturbation theory.
As explained in section \ref{schiral}, the chiral Lagrangian itself
takes the same form for any number of flavors.
To transcribe the operators, however, is more complicated.
It turns out that, at the order I am working, one needs the following set
\begin{eqnarray}
\co_1 &=& f^4 (\Sigma\del_\mu \Sigma^{\dag})_{d's'}
		(\Sigma\del_\mu \Sigma^{\dag})_{ds} \\
\co_2 &=& f^4 (\Sigma\del_\mu \Sigma^{\dag})_{d's}
		(\Sigma\del_\mu \Sigma^{\dag})_{ds'} \\
\co_3 &=& - 2 \mu^2 f^4 \Sigma_{d's} \Sigma^{\dag}_{ds'}   \\
\co_4 &=& f^4 (\del_\mu \Sigma_{d's}) (\del_\mu \Sigma^{\dag}_{ds'})  \\
\co_5 &=& f^4 	(\Sigma\del_\mu \Sigma^{\dag})_{d's'}
		(\Sigma^{\dag}\del_\mu \Sigma)_{ds} \ .
\end{eqnarray}
The overall factors are chosen for later convenience.

To transcribe the quark operators, they must be written in terms of
representations of the $SU(N)_L\times SU(N)_R$ chiral group.
The first step is to rewrite the operators
in terms of left- and right-handed bilinears
\begin{equation}
\label{vvaadecomp}
[V'V]_i = \half ([L' L]_i + [L' R]_i) + \co^-\ ,\quad
[A'A]_i = \half ([L' L]_i - [L' R]_i) + \co'^-\ ,
\end{equation}
where $i=1,2$ labels the number of color traces.
The negative parity operators $\co^-$ and $\co'^-$ do not contribute
to the matrix elements under study and can be ignored.
I am using the compact notation
\begin{eqnarray}
 [V'V]_{1} &=& [\bar s'_a \g\mu d'_b]    \ [\bar s_b \g\mu d_a]  \ ,\\ {}
 [A'A]_{1} &=& [\bar s'_a \g\mu \g5 d'_b] \ [\bar s_b \g\mu \g5 d_a] \ ,\\ {}
 [L'L]_{1} &=& [\bar s'_a \g\mu \pl d'_b] \ [\bar s_b \g\mu \pl d_a] \ ,\\ {}
 [L'R]_{1} &=& [\bar s'_a \g\mu \pl d'_b] \ [\bar s_b \g\mu \pr d_a] \ ,
\end{eqnarray}
with similar definitions for operators with two color traces.

The left-handed bilinears $L$ and $L'$ transform as $({\bf (N^2\!-\!1)},1)$,
i.e. as adjoints under $SU(N)_L$ and singlets under $SU(N)_R$.
Under $SU(N)_L$, $[L'L]$ lies in the symmetric product of two adjoints,
\begin{equation}
({\bf (N^2\!-\!1)} \otimes {\bf (N^2\!-\!1)})_{\rm symm}
= {\bf 1} + {\bf (N^2\!-\!1)} + {\bf \cs} + {\bf \ca} \ .
\end{equation}
The representations $\cs$ and $\ca$ are both traceless, and are
respectively symmetric and antisymmetric under interchange of quark indices.
%%%%%%%%%%%%%%%%%%%%%%%%%%%%%%%%%%%%%%%%%%%%%%%%%%%%%%%%%%%%%%%%%
%this is the generalization of the $su(4)$ decomposition
%$$({\bf 15} \times {\bf 15} )_{\rm symm}
%= {\bf 1} + {\bf 15} + {\bf 84}(\ytabi) + {\bf 20}(\ytabc).$$
%%%%%%%%%%%%%%%%%%%%%%%%%%%%%%%%%%%%%%%%%%%%%%%%%%%%%%%%%%%%%%%%
They have dimensions $(N-1)N^2(N+3)/4$ and $(N-3)N^2(N+1)/4$ respectively.
Since $L'$ and $L$ have no flavors in common, their product
must be a combination of ${\bf \cs}$ and ${\bf \ca}$.
The combinations belonging to one or other representation are
\begin{equation}
 [L'L]_2 \pm [L'L]_1 =
[\bar s'_a \g\mu \pl d'_a] \ [\bar s_b \g\mu \pl d_b]
\pm [\bar s'_a \g\mu \pl d_a] \ [\bar s_b \g\mu \pl d'_b]
\in \left\{ \begin{array}{l} {\bf \cs}\\ {\bf \ca} \end{array} \right. .
\end{equation}
In chiral perturbation theory the lowest dimension operators transforming
in these representations are
\begin{eqnarray*}
({\bf \cs},{\bf 1}): && \co_\cs = \co_1 + \co_2 \ ; \\
({\bf \ca},{\bf 1}): && \co_\ca = \co_1 - \co_2 \ .
\end{eqnarray*}
Thus the transcription of the continuum operators is
\begin{eqnarray}
[L'L]_2 + [L'L]_1 &=& \alpha_S \co_\cs + O(p^4)\ , \\ {}
[L'L]_2 - [L'L]_1 &=& \alpha_A \co_\ca + O(p^4)\ ,
\end{eqnarray}
where $\alpha_S$ and $\alpha_A$ are new, unknown coefficients.
One can estimate their values by taking matrix elements
between a $K$ and a $\bar{K'}$,
using the vacuum saturation approximation for the l.h.s.,
and tree level chiral perturbation theory for the r.h.s.
The resulting estimates are $\alpha_S=\frac43$ and $\alpha_A=\frac23$.
The same results are obtained using external $\bar d's$ and $\bar s'd$ kaons.

The $[L'R]$ operators transform as $({\bf(N^2\!-\!1)},{\bf(N^2\!-\!1)})$.
They can thus be written as linear combinations of the three operators
$\co_{3-5}$, which also transform in this way
\begin{equation}
\label{lrtrans}
[L'R]_i = \beta_i \co_3 + \gamma_i \co_4 + \delta_i \co_5 + O(p^4) \ .
\end{equation}
There are six unknown coefficients, three for each color contraction.
To estimate their size I first take the matrix element
between a $K$ and a $\bar{K'}$.
At tree level, only $\co_5$ has a non-vanishing matrix element.
Using vacuum saturation for the matrix element of $[L'R]_i$ yields
the estimates $\delta_1=\frac13$, $\delta_2=1$.

To estimate the $\beta_i$ I take the matrix element between
$\bar d's$ and $\bar s'd$ kaons.
In chiral perturbation theory one finds, at tree level,
\begin{equation}
\label{lrtree}
\vev{K(\bar d's)|[L'R]_i|K(\bar s'd)}
= -4 \beta_i \mu^2 f^2 + O(\gamma_i m_K^2) \ .
\end{equation}
To use the vacuum saturation approximation I first Fierz transform the
operators, e.g.
\begin{equation}
[L'R]_1 = - 2 [\bar s'_a \pr d_a] \ [\bar s_b \pl d'_b] \ ,
\end{equation}
and then use
\begin{equation}
\vev{0|\bar s_b \g5 d'_b |K(\bar s'd)} = i \mu \sqrt2 f_K \ ,
\end{equation}
to give
\begin{eqnarray}
\label{lrvacsat}
\vev{K(\bar d's)|[L'R]_1|K(\bar s'd)}_{\rm vac} = -4 \mu^2 f^2 \ , \\
\vev{K(\bar d's)|[L'R]_2|K(\bar s'd)}_{\rm vac} = -\frac43 \mu^2 f^2 \ .
\end{eqnarray}
The resulting estimates are $\beta_1=1$ and $\beta_2=\frac13$.
This method gives no estimate for the $\gamma_i$.

In summary, each quark operator can be written as
$\sum_j c_j \co_j$ with corrections of $O(p^4)$.
The coefficients are collected in Table \ref{tabone},
where I use the definitions
\begin{equation}
\alpha_\pm=\half(\alpha_S \pm \alpha_A)\ .
\end{equation}
In vacuum saturation approximation these are
$\alpha_+=1$ and $\alpha_-=\frac13$.

\begin{table}
% tabbing stuff
\begin{center}
\begin{tabular}{cccccc}
Operator   & $c_1$	& $c_2$	     & $c_3$	 & $c_4$      & $c_5$ \\ \hline
$2 [V'V]_1$& $\alpha_-$ & $\alpha_+$ & $\beta_1$ & $\gamma_1$ & $\delta_1$ \\
$2 [V'V]_2$& $\alpha_+$ & $\alpha_-$ & $\beta_2$ & $\gamma_2$ & $\delta_2$ \\
$2 [A'A]_1$& $\alpha_-$ & $\alpha_+$ &$-\beta_1$ &$-\gamma_1$ &$-\delta_1$ \\
$2 [A'A]_2$& $\alpha_+$ & $\alpha_-$ &$-\beta_2$ &$-\gamma_2$ &$-\delta_2$ \\
\end{tabular}
\end{center}
\caption[tabone]
{Decomposition of quark operators into chiral operators $\sum_j c_j \co_j$.}
\label{tabone}
\end{table}

All four operators in Eqs. (\ref{v1defeqn})-(\ref{a2defeqn})
contain ``left-left'' and ``left-right'' parts.
Equation (\ref{lrtree}) shows that the matrix elements
of ``left-right'' operators are generically of $O(1)$,
unlike those of ``left-left'' operators which are of $O(p^2)$.
Nevertheless, at tree level, and for the external flavors $K$ and $K'$,
all four operators have matrix elements of $O(p^2)$.
The choice of external flavors has removed the $O(1)$ contributions.
As we will see, however, these contributions
do have an effect when loops are included.

The calculation of the chiral logs parallels that for $B_K$.
The same diagrams contribute and the general result is
\begin{eqnarray}
\nonumber   \vev{\bar{K'^0}|c_j \co_j|K^0} &=&
(c_1\!-\!c_5)\ 2 m_K^2 f_K^2  \\
\nonumber
&+& c_2\ f^2 \left( 2 m_K^2 I_2(m_K)
 - (m_K^2+m_{ss}^2)I_1(m_{ss}) - (m_K^2+m_{dd}^2)I_1(m_{dd}) \right) \\
\nonumber
&+& c_3\ 2 \mu^2 f^2
\left( I_1(m_{ss})+I_1(m_{dd})-2I_1(m_K)-2I_2(m_K) \right) \\
\nonumber
&+& c_4\ f^2 \left(4m_K^2 I_1(m_K) - 2m_K^2 I_2(m_K)\right. \\
\label{generallogs}
&&\qquad
\left. - (m_K^2+m_{ss}^2)I_1(m_{ss}) - (m_K^2+m_{dd}^2)I_1(m_{dd}) \right)\ .
\end{eqnarray}
I have simplified the result using $m_{s'}=m_s$ and $m_{d'}=m_d$.
Only the operators $\co_1$ and $\co_5$ contribute to
Figs. \ref{fsix}(a) and (b),
and the corrections simply convert $f^2$ into the 1-loop result $f_K^2$.
The general expression for $f_K$ at 1-loop is given in Section \ref{sfpi}.
These are the only contributions of $\co_1$ and $\co_5$.
The remaining three operators contribute only to Figs. \ref{fsix}(c)-(e).
The pions in the loops in these diagrams always consist of a primed quark
and an unprimed antiquark (or vica-versa). These states are automatically
mass eigenstates, for any number of flavors,
so the result is valid for all $N\ge4$.

There are no quartic divergences in Eq. (\ref{generallogs}).
The only quadratic divergences at this order are
proportional to $c_2 m_K^2 \Lambda^2$.
These can be absorbed into the coefficient $c_1$,
and correspond to mixing between $\co_1$ and $\co_2$.
The quadratic divergences proportional to $c_3$ cancel;
this is essential, because they are not proportional to $m_K^2$,
and could not be absorbed by a redefinition of the coefficients.
The quadratic divergences proportional to $c_4$ cancel because
$2m_K^2 = m_{ss}^2 + m_{dd}^2$. This is not required by chiral symmetry,
however, as a divergence could have been absorbed into $c_5$.
Finally, there are also quadratic divergences hidden in $f_K^2$,
which can be absorbed into $f$, as discussed previously.

I now collect the results for the $B$-parameters,
using the general formula Eq. (\ref{generallogs}),
the decompositions given in Table \ref{tabone},
and the definitions Eqs. (\ref{ovdefeqn})-(\ref{bvdefeqn}).
I present the results for $m_s=m_d$,
since it turns out that these also apply in the quenched approximation.
The generalizations to non-degenerate quarks are easy to obtain
from the above.
For the vector $B$-parameters I find
\begin{eqnarray}
\label{bv1result}
B_{V1} &=& \frac38(\alpha_- - \delta_1)
	+\frac38\alpha_+(I_2-2I_1)
	-\frac34 \beta_1 {\mu^2 \over m_K^2} I_2
	-\frac38 \gamma_1 I_2 \ , \\
\label{bv2result}
B_{V2} &=& \frac38(\alpha_+ - \delta_2)
	+\frac38\alpha_-(I_2-2I_1)
	-\frac34 \beta_2 {\mu^2 \over m_K^2} I_2
	-\frac38 \gamma_2 I_2 \ .
\end{eqnarray}
Since $m_K=m_{ss}=m_{dd}$, the arguments of the $I_i$ are all
the same, and I have dropped them.
The results for $B_A$ are the same except that terms proportional
to $\beta_i$, $\gamma_i$ and $\delta_i$ have the opposite sign.
A check on these results is that they combine to give the same result
for $B_K=B_{V1}+B_{V2}+B_{A1}+B_{A2}$ as found previously
(Eq. (\ref{bkdegenerate}))
\begin{equation}
B_K = \frac34 \alpha_S (1 + I_2-2I_1) \ ,
\end{equation}
providing one identifies $B=\frac34 \alpha_S$.

The corrections to $B_V$ and $B_A$ include a term of the form
$I_2/m_K^2$ which is proportional to $\ln(m_K)$.
This is enhanced by a factor of $\mu^2/m_K^2$ over the usual chiral logs.
The general possibility of enhanced chiral logs was noticed long ago
by Pagels and Langacker \cite{langackerpagels}.
The enhanced log here is the remnant of the fact that
$\co_3$ is of $O(1)$ in the chiral expansion, rather than of $O(p^2)$.
For small enough $m_K$ the enhanced log dominates over the
``leading order'' terms, and it diverges in the chiral limit.
The vector-vector and axial-axial matrix elements do not themselves diverge,
but contain a chiral log proportional to $m_K^2 \ln(m_K)$,
rather than the usual $m_K^4 \ln(m_K)$.
The unknown scale in the log can be absorbed into $\delta_i$,
and corresponds to mixing between $\co_3$ and $\co_5$.

These results can be adapted to the quenched approximation
with the by now standard procedure.
The reasoning is identical to that for $B_K$,
because the quark diagrams are the same.
The only diagrams containing internal quark loops are those that change
$f^2$ to $f_K^2$, and these cancel in $B-$parameters.
The other diagrams, those of Fig. \ref{fseven},
are included in the quenched approximation.
There are only two possible differences between quenched and full theories.
The first comes from the fact that the mass eigenstates are not the same
for flavor diagonal states.
This is not relevant here, however, since
all pions in the loops are flavor off-diagonal.
The second comes from $\eta'$ loops, which do not contribute to the full
theory, but might be present in the quenched approximation.
The quark diagram is that shown in Fig. \ref{fseven}(e).
As for $B_K$, however, this diagram does not contribute if $m_s=m_d$.
For $\co_1$, $\co_2$ and $\co_5$, this is because the $\eta'$ does not
couple to these operators. For $\co_3$ and $\co_4$, it is because
the $\eta'$, although it couples to the operators, cannot appear in a loop
like that in Fig. \ref{fseven}(a) because the flavor indices are wrong.
The $\eta'$ does appear in two loop diagrams.
The conclusion is that, for $m_s=m_d$, the results of Eqs.
(\ref{generallogs})-(\ref{bv2result}) apply without change
to the quenched approximation.
For $m_s\ne m_d$, however, there will be additional contributions from
Fig. \ref{fseven}(e).

%%%%%%%%%%%%%%%%%%%%%%%%%%%%%%%%%%%%%%%%%%%%%%%%%%%%%%%%%%%%%%%%%%%%%%%%%
%I close this Section with a remark about the quark diagrams in the
%$N\ge4$ flavor theory, which illustrates how the diagrammatic analysis works.
%Since all quarks have different flavors,
%the breakup of the four-fermion operator into two bilinears is well defined.
%Recall that, at leading order, and for degenerate masses,
%only Figs. \ref{fseven}(a), (c) and (d) contribute.
%In these diagrams the bilinears must have flavors $\bar sd'$ and $\bar s'd$,
%since the quark and antiquark emerging from each bilinear
%end up in different external kaons.
%This implies that only the operators $\co_{3-5}$ can contribute to these
%diagrams, for only these operators can produce pions of the required flavors.
%This is indeed what was found above.
%%%%%%%%%%%%%%%%%%%%%%%%%%%%%%%%%%%%%%%%%%%%%%%%%%%%%%%%%%%%%%%%%%%%%%%

There is one complication in the numerical calculation of the $B-$parameters
which arises if one uses staggered fermions.
I describe this briefly here; for more details see, for example, Refs.
\cite{toolkit,book}.
Each lattice staggered fermion represents four continuum flavors,
so the quark fields have an additional flavor index.
The bilinears that are actually used have these extra indices contracted as in
\begin{equation}
\bar s \g\mu d \longrightarrow
\bar s_1 \g\mu d_1 + \bar s_2 \g\mu d_2
- \bar s_3 \g\mu d_3 - \bar s_4 \g\mu d_4 \ ,
\end{equation}
where the color indices are implicit.
This particular flavor is that of the lattice Goldstone pion,
and in numerical calculations the external kaons have this flavor.
The analysis given above is valid for any number of flavors,
and thus applies directly to these generalized operators.
One simply has to break each matrix element up into parts,
each of which is like those in Eqs. (\ref{v1defeqn})-(\ref{a2defeqn}).
Keeping track of flavor factors, one finds that the form of
the final result is unchanged. The only difference is that most of the pions
in the loops have flavors such as $\bar s'_1 d_2$,
and thus are not Goldstone pions.
In fact, it is straightforward to show that only one out of sixteen pions
is a Goldstone pion.
This is important at finite lattice spacing,
because non-Goldstone pions are slightly more massive than the Goldstone pion,
and one must use the appropriate mass in the loop integrals.
In addition, the vertices of the non-Goldstone pions will differ
{}from those of the Goldstone pions by terms of $O(a)$.
The effect of this is to make the replacement
\begin{equation}
\label{staggprescription}
I_2(m_K)/m_K^2 \longrightarrow \frac1{16} I_2(m_G)/m_G^2 +
\frac{15}{16} R_{NG} I_2(m_{NG})/m_{NG}^2 \ ,
\end{equation}
where $m_G$ and $m_{NG}$ are, respectively,
the mass of the Goldstone and non-Goldstone pions,
and $R_{NG}$ is the ratio of non-Goldstone to Goldstone coefficients.
(I am assuming here that all the non-Goldstone pions are degenerate,
which, numerically, is a good approximation.)
Similar replacements should be made in all the loop integrals in Eqs.
(\ref{bv1result}) and (\ref{bv2result}).

\section{$\eta'$ LOOPS}
\label{seta'}
%%% sec5.tex
%% file: sec5.tex
%%\section{$\eta'$ LOOPS}
%%\label{seta'}

The major new feature of quenched chiral perturbation theory
is the appearance of the light $\eta'$ in loops.
Thus far I have considered only quantities
which do not have $\eta'$ contributions at one-loop order.
Clearly it would be preferable to understand such contributions,
since this would increase the number of quantities whose finite volume
dependence could be predicted.
In addition, it is potentially dangerous to leave an unsettled problem
in one part of the theory, as it may feed back at higher loop order into the
quantities I have studied.

I consider the effects of $\eta'$ loops in theory with
$N$ degenerate valence quarks.
Using degenerate quarks simplifies the discussion in two ways.
First, the $\eta'$ can be chosen as one of the mass eigenstates,
which allows the effects of $\eta'$ loops to be clearly distinguished.
Second, as discussed above, the $\eta'$ only couples to the pions
through the mass term in the chiral Lagrangian, $\cl_m$.
This means, in particular, that $\eta'$ loops only affect mass renormalization,
and I focus on this in the following discussion.

\begin{figure}
\vspace{2.1truein}
%\special{psfile=fig8.ps hoffset=12 voffset=12 hscale=0.9 vscale=0.9}
% for uw
\includegraphics{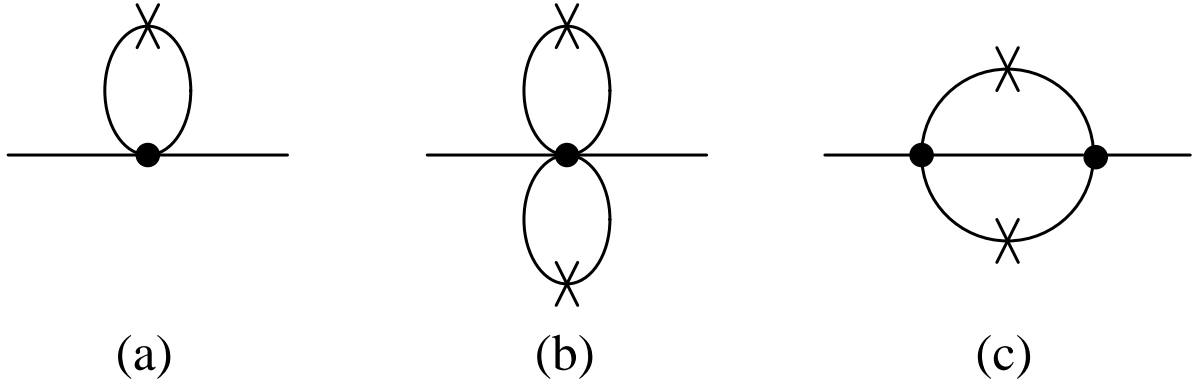}
%%\makebox[\textwidth][c]{(a) \hskip 3truein (b)}
\caption[feight]{\fc
Diagrams contributing to mass renormalization in the quenched
approximation. The circle represents the vertex from the mass term
in the chiral Lagrangian, the cross an insertion of the gluonic
amplitude ${\cal L}_{\eta'}$. Lines with a cross are $\eta'$ propagators;
the others can be either pion or $\eta'$ propagators.}
\label{feight}
\end{figure}

In the quenched approximation, the only diagrams which renormalize the
pion mass at leading order are those of Figs. \ref{fthree}(d)-(f).
Note that only the $\eta'$ appears in these loops,
since non-singlet states cannot annihilate into gluons.
As discussed at the end of Section \ref{schiral},
the contribution from Fig. \ref{fthree}(f),
which arises from the vertex $V_2'(0) \eta' \Tr(M\eta'\pi^2)$,
can be rotated away by a field redefinition.
Its effect appears in a change in the value of the constant $A$
in $\cl_{\eta'}$ (Eq. (\ref{eta'twopoint})).
Thus I need only consider Figs. \ref{fthree}(d) and (e).
These are represented in chiral perturbation theory by
Fig. \ref{feight}(a), where the cross is the gluon annihilation vertex,
$\cl_{\eta'}$.
The renormalized mass at one-loop from Fig \ref{feight}(a) is
\begin{eqnarray}
\label{mpirenorm}
m_\pi^2 &=& 2 \mu m \left(1 + X({\rm Fig. 8a}) \right) \\
X({\rm Fig. 8a}) &=&
{1\over N f^2} \int_k G(k,m_{\eta'})^2 (A m_0^2 + (A-1) k^2)\\
\label{Xfig8ares}
&=& {{1}\over{N}} \left(A {m_0^2\over m_{\eta'}^2} - (A-1) \right)
I_2(m_{\eta'}) + {{1}\over{N}} (A-1) I_1(m_{\eta'})
\end{eqnarray}
The mass in the loop is $m_\eta'$, which equals $m_\pi$ at tree level.
This result agrees with Ref. \cite{bernardgolterman}.
The $k^2$ term in the integrand leads to a correction of the usual type:
it is proportional to $m_{\eta'}^2 \ln(m_{\eta'})$,
and vanishes in the chiral limit.
The $m_0^2$ term, by contrast, leads to a correction proportional to
$m_0^2 \ln(m_{\eta'})$ which diverges in the chiral limit.
This enhanced chiral log is similar those appearing in $B_V$ and $B_A$.
The additional dimensionful parameter $m_0$ violates the standard power
counting rules of chiral perturbation theory, according to which each
loop comes with a factor of $m_\pi^2$, possibly multiplied by logarithms.
For small enough $m_\pi$, the one-loop correction proportional to
$m_0^2\ln(m_{\eta'})$ is not small, and one must sum all such terms.

%%% Fig 8

In practice, as stressed in Ref. \cite{bernardgolterman},
this summation may not be numerical important.
This is because in present lattice calculations the pions are not very light,
$m_\pi\ge 400$ MeV.
To estimate the size of the correction, I extract the leading logarithm
\begin{equation}
X({\rm Fig. 8a}) = - \delta \ln\left({m_{\eta'}^2\over\Lambda^2}\right) \ .
\end{equation}
Here I have introduced the parameter
\begin{equation}
\delta = {A m_0^2 \over N (4 \pi f)^2} \approx 0.2 \ ,
\end{equation}
where the estimate uses $N=3$, $A=1$, and $m_0=900$ MeV.
Taking $\Lambda\approx m_\rho$,
the magnitude of the logarithm is roughly unity for $m_\pi=400$ MeV.
Thus the correction increases $m_\pi^2$ by only $20\%$,
and resummation is not required.
Since the logarithm is not large, however, it does not dominate over
constants and power corrections,
and the full expression of Eq. (\ref{Xfig8ares}) should be used.

Bernard and Golterman have begun a program of calculations,
in which they include $\eta'$ loops only at one-loop order,
for the general case of non-degenerate quarks \cite{bernardgolterman}.
Their predictions can be compared to numerical results for various
quantities. My interest in this section is not phenomenological,
but rather to understand the apparent divergence of the correction in
the chiral limit.
To address this question one must resum the leading logarithms
of the form $m_0^{2n} \ln(m_{\eta'})^n/f^{2n}$.
This is relatively straightforward,
because the leading logs come only from diagrams in which ${\cal L}_{\eta'}$
appears on a propagator both ends of which come from the same vertex.
An example is shown in Fig. \ref{feight}(b).
Diagrams involving $\eta'$ propagators connecting two or more vertices
have fewer logs. For example, the diagram of Fig. \ref{feight}(c)
gives a correction proportional to $m_0^4 m_{\eta'}^2/f^6$,
with no logs at all.

\begin{figure}
\vspace{2.0truein}
%\special{psfile=fig9.ps hoffset=36 voffset=12 hscale=0.9 vscale=0.9}
% for uw
\includegraphics{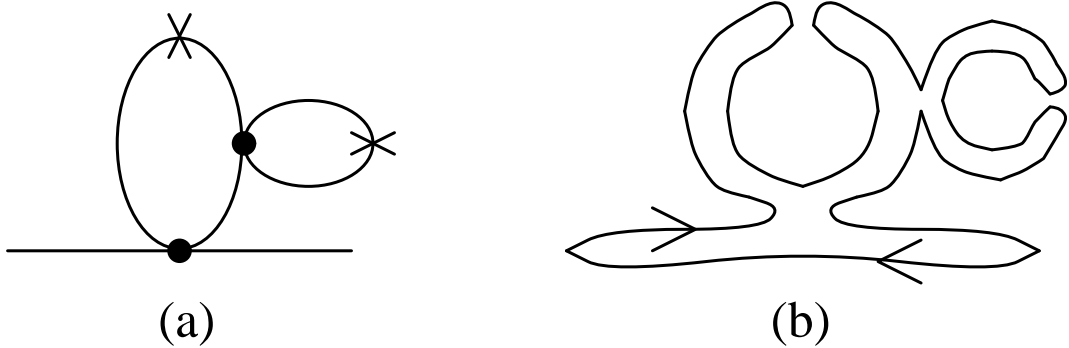}
%%\makebox[\textwidth][c]{(a) \hskip 3truein (b)}
\caption[fnine]{\fc
An example of a diagram corresponding to self-consistent leading-log
evaluation of $m_\pi$ in the quenched approximation:
(a) in chiral perturbation theory; (b) in terms of quarks.
Notation as in Fig. 8.}
\label{fnine}
\end{figure}

The bubbles exemplified by Figs. \ref{feight}(a) and (b)
sum up to an exponential
\begin{equation}
\label{Xfig8aresum}
m_\pi^2 = 2 \mu m\exp\left(X({\rm Fig. 8a})\right) \ ,
\end{equation}
the leading logarithmic part of which is
\begin{equation}
\label{mpiresum}
m_\pi^2 = 2 \mu m \left({\Lambda^2 \over m_{\eta'}^2}\right)^{\delta} \ .
\end{equation}
%Here I have distinguished between the pion and $\eta'$ masses,
%since the former appears only on the external legs, the latter in the loops.
It is straightforward to show that the same exponential factor
corrects any vertex coming from ${\cal L}_m$,
so that the entire mass term is replaced by
\begin{equation}
\label{newLm}
{\cal L}'_m = \left({\Lambda^2 \over m_{\eta'}^2}\right)^{\delta} {\cal L}_m\ ,
\end{equation}
with the proviso that no $\eta'$ fields coming from this vertex
be joined together by an insertion of ${\cal L}_{\eta'}$.
%This result follows from standard trigonometric identities involving
%sines and cosines.
One corollary of this result is that the $\eta'$ mass itself
is corrected in the same way as $m_\pi$, so that they remain equal.
This means that Eq. (\ref{mpiresum}) should be interpreted as a
self-consistent equation for the common pion and $\eta'$ mass,
with $m_{\eta'}$ on the r.h.s. replaced by $m_\pi$.
Solving this equation yields the final result
\begin{equation}
\label{mpifinal}
m_\pi^2 = \left(2\mu m\right)^{1\over1+\delta}
	\left(\Lambda^2\right)^{\delta\over1+\delta} \ .
\end{equation}
Solving Eq. (\ref{mpiresum}) self-consistently
corresponds to summing all the diagrams of the form Fig. \ref{fnine}(a),
in which each internal $\eta'$ propagator is decorated with any number
of bubble sums.
In terms of quarks, these ``cactus'' diagrams look like Fig. \ref{fnine}(b).
%All these diagrams are present in the quenched approximation.

%%% Fig 9

The result Eq. (\ref{mpifinal}) seems very odd at first sight.
For one thing, $m_\pi^2$ has a non-analytic dependence on the quark mass.
For another, it appears to contradict the strong numerical evidence that
$m_\pi^2 \propto m$ in quenched as well as full theories \cite{hmks}.
To investigate further, I substitute Eq. (\ref{mpifinal})
back into Eq. (\ref{newLm}) to give (using also Eq. (\ref{massterm}))
\begin{equation}
\label{finalLm}
{\cal L}'_m = -{f^2 \over 4}
	\left(2\mu m\right)^{1\over1+\delta}
	\left(\Lambda^2\right)^{\delta\over1+\delta}
\left( \cos(\phi_0) \Tr\big[\Sigma\!+\!\Sigma^{\dag}\big] +
 i \sin(\phi_0) \Tr\big[\Sigma\!-\!\Sigma^{\dag}\big] \right) \ .
\end{equation}
This says that, at leading-log order, the effect of $\eta'$ loops
amounts to a change in the coefficient multiplying the mass term.
The quenched chiral Lagrangian looks the same as that in the full theory
\footnote{%
%%%%%%%%%%%%%%%%%%%%%%%%%%%%%%%%%%%%%%%%%%%%%%%%%%%%%%%%%%%%%%%%%%%%%
Of course, in the application of this Lagrangian, one must follow
the rules discussed in the sections 3 and 4 for
discarding diagrams containing internal quark loops.}
%%%%%%%%%%%%%%%%%%%%%%%%%%%%%%%%%%%%%%%%%%%%%%%%%%%%%%%%%%%%%%%%%%%%%
as long as one uses a new mass $m'$
\begin{equation}
m' = m \left({\Lambda^2\over 2\mu m}\right)^{\delta\over1+\delta} \ .
\end{equation}
Looked at this way, the result is less strange.
All the relationships between amplitudes involving different
numbers of pions remain valid, e.g. the Weinberg scattering length formulae.
These relations involve only the physical parameters $f_\pi$ and
$m_\pi$, and not the quark masses.
In particular, there is no contradiction with exact lattice Ward identities
that can be derived with staggered fermions \cite{toolkit,pipinew}.

Associated with the enhanced chiral logarithms is an
``enhanced'' finite volume dependence.
The precise form of this dependence is discussed in section 7.
The important point for present purposes is that it too
is as a function of physical quantities alone, and not of $m$ (or $m'$).
Thus one cannot use the finite volume behavior to observe the
non-analytic dependence on $m$.

Is there any way of observing this dependence?
The remaining hope rests with the numerical lattice results for the quark
mass dependence of $m_\pi^2$ and the condensate $\vev{\bar\psi\psi}$.
The predicted dependence of $m_\pi^2$ on $m$ is given in Eq. (\ref{mpifinal}).
That of the condensate may be deduced from Eq. (\ref{finalLm})
(or from the formula of Gell-Mann, Oakes and Renner)
\begin{equation}
\vev{\bar\psi\psi} \propto m^{-\delta\over1+\delta} \ .
\end{equation}
These predictions are in apparent conflict with numerical simulations
which find that $m_\pi^2 \propto m_q$
and $\vev{\bar\psi\psi} =\vev{\bar\psi\psi}_0 + O(m_q)$,
where $m_q$ is the bare lattice quark mass \cite{hmks}.
The results are probably good enough to rule out $\delta\approx0.2$,
{\em assuming that we identify $m_q$ with $m$}.

In full QCD this identification is needed for the following reason.
The bare quark mass $m_q$ is part of a scalar source $s$.
Functional derivatives with respect to $s$
yield correlators involving scalar bilinears in the presence of the source.
Setting the source to zero we obtain the correlators in the massless theory,
the condensate being the simplest example.
We expect these correlators to exist in a regulated theory.
For this to be so, the chiral Lagrangian must contain only
positive integral powers of $s$.
This means that, up to an overall constant, and up to corrections
of $O(m^2)$, we should identify $m_q$ with $m$.

In quenched QCD, by contrast, one does not have a Lagrangian framework,
and I see no reason why the connection between $m$ and $m_q$ should be
analytic.
I propose that $m_q$ should be identified with $m'$
(up to an overall factor), rather than with $m$.
This would imply the relationship
\begin{equation}
m_q \propto m^{1+\delta} \ .
\end{equation}
As far as I can tell, once this identification is made
there are no observable effects of the non-analytic dependence on $m$.

\section{FINITE VOLUME DEPENDENCE}
\label{sfinite}
%%% sec6.tex
%% file: sec6.tex
%%\section{FINITE VOLUME DEPENDENCE}
%%\label{sfinite}

The discussion thus far concerns quantities calculated in infinite volume.
Numerical simulations, by contrast, are done in finite boxes.
These have finite extent in both space and in Euclidean ``time''.
I denote the length of the lattice in the spatial directions by $L_i$,
$i=1\!-\!3$, and the length in the time direction by $L_4$.
The latter is related to the temperature by $T=1/L_4$.
In the following discussion, I assume that $L_4>>L_i$,
so that the dominant finite size effect is due to the spatial volume.
%I will often just refer to this as the ``volume''.

Gasser and Leutwyler have shown that the same loop graphs which
give rise to the leading chiral logarithms also allow one to predict the
leading finite volume dependence \cite{condfinitev}.
They have presented results for $f_\pi$ and $m_\pi$ in a
theory with $N$ degenerate flavors. It is simple to extend their
method to any quantity for which the chiral logs are known.
The only subtlety concerns the applicability of their method to
the quenched approximation.

Their method depends upon the result that the volume dependence
in chiral perturbation theory enters only through the pion propagators.
The chiral Lagrangian itself does not depend on the volume,
and so the vertices in chiral perturbation theory are unchanged.
This is also true of external sources, such as that representing $\co_K$.
Thus the constants $f$, $\mu$, and $B$ are volume independent.
The volume dependence of the pion propagator enters through the boundary
conditions, which are periodic in all four Euclidean directions.
This means that the integral over momenta is replaced by a sum,
in which $p_\mu$ is an integer multiple of $2\pi/L_\mu$.
In position space, the propagator is a sum over periodic images
\begin{equation}
\label{finiteVpion}
G_L(x,m) = G(x,m) + \sum_{n_\mu\ne 0} G(x +n_\mu L_\mu,m) \ ,
\end{equation}
where $n_\mu=(1,0,0,0)$, $(-1,0,0,0)$, $(0,1,0,0)$, etc.
Volume dependence comes from replacing
$G$ with $G_L$ in loop integrals such as $I_1$ and $I_2$.

The volume independence of the chiral Lagrangian is clearly crucial to
the method. The arguments for this are presented
in detail in Ref. \cite{glfiniteL}.
The ingredients are\footnote{%
In addition, one must work at volumes large enough that chiral symmetry
is not restored.}
%%%%%%%%%%%%%%%%%%%%%%%%%%%%%%%%%%%%%%%%%%%%%%%%%%%%%%%%%%%%%%%%%%%%%%%%%%
(a) that chiral Ward identities remain valid in finite volume,
(b) causality,
(c) that the pion propagator at fixed distance should approach the
    infinite volume form as the volume increases to infinity, and
(d) the permutation symmetry between the four directions $\mu=1-4$.
The first three ingredients establish the existence of a finite volume
chiral Lagrangian, and that the coefficients of its leading terms
are the same as at infinite volume.
Volume dependence suppressed by powers of $1/L^2$ is allowed,
so there can be terms such as
$\sum_i\Tr(\del_i \Sigma \del_i \Sigma^{\dag})/L_i^2$.
Such terms are ruled out by the fourth ingredient, permutation symmetry.
This symmetry implies that,
if there is any dependence on the spatial lengths $L_i$,
there must also be a dependence on $L_4$, and thus on $T$.
But this contradicts a standard field theoretic result, namely
that the temperature enters the functional integral representation
of the partition function only through the boundary conditions.
Thus permutation symmetry rules out $L_i$ dependence in the vertices
at any order in the chiral expansion.
For the symmetry to be present, the boundary conditions must
be the same in all four directions. In particular, since the finite
temperature functional integral has antiperiodic boundary conditions on
fermions in the time direction, the boundary conditions must also be
antiperiodic in the spatial directions.

In the quenched approximation only the first and third ingredients are
present. Concerning the former,
one can derive the Ward identities on the lattice independent
of the quenched approximation \cite{wilsonWI,toolkit}.
As for the latter, it is reasonable to assume that the boundaries
become unimportant as they recede to infinity.
But the lack of dynamical quarks means that one cannot even
formulate the second ingredient, causality.
Causality requires that the anticommutator of fermion fields
vanish outside the forward lightcone.
In the quenched approximation, one cannot construct the transfer matrix
and obtain an operator formulation in which quarks are represented
by anticommuting operators.
A less formal way of seeing this is to note that,
without quark loops, there is no Pauli exclusion principle,
so that, in some sense, fermion fields no longer anticommute.
The lack of quark loops also means that the fourth ingredient,
permutation symmetry, is not relevant.
The requirement of antiperiodic boundary conditions applies to
internal quark loops,
for it is these loops which build up the contribution of states
containing quarks to the partition function.
Since these loops are absent, one is free to choose the boundary
conditions on the valence quarks as one pleases.

The absence of two of the four ingredients means that that one cannot
rigorously argue for the volume independence of the chiral Lagrangian
in the quenched approximation.
In fact, as already discussed, even the existence of such a Lagrangian is
an assumption. I will simply extend this assumption by taking the
quenched chiral Lagrangian to have volume independent coefficients.
I think this is reasonable, and in the following I
give a qualitative argument in its support.
Ultimately, however, it must be tested by numerical simulations.

The physical origin of the finite size effects from chiral loops
is that particles are surrounded by a cloud of virtual pions.
The pions in this cloud can propagate ``around the world'' due to
the periodicity, the propagator falling roughly like $\exp(-m_\pi L_i)$.
If there were no interactions, then there would be no pion cloud,
and the only effect of finite volume would be to restrict the momenta
to discrete values.
In particular, masses would remain unchanged.
L\"uscher has given a general analysis of finite volume effects
arising from polarization clouds
on the masses of stable particles \cite{luscherexp}.
For the pion, his analysis implies that the leading order effect
comes from a ``tadpole'' diagram, similar to that of Fig. \ref{fone}(b),
except that the vertex and propagator are fully dressed.
The crucial result is that volume dependence enters only through the pion
propagator,
which goes ``around the world'' in each of the six spatial directions.
L\"uscher's general formula simplifies when applied to pions,
because one can do a chiral expansion of the propagator and vertices.%
%%%%%%%%%%%%%%%%%%%%%%%%%%%%%%%%%%%%%%%%%%%%%%%%%%%%%%%%%%%%%%%%%%%%%%%%%
\footnote{%
The reader consulting Ref. \cite{luscherexp} might be misled by the fact
that the leading term in L\"uscher's formula falls like $\exp(-\sqrt{3/4}mL)$,
i.e. more rapidly than $\exp(-mL)$. This term, however, is absent for pions
because it is proportional to the three pion coupling, which is zero.
The dominant term falls as $\exp(-mL)$,
and the leading corrections are proportional to $\exp(-\sqrt2mL)$.}
%%%%%%%%%%%%%%%%%%%%%%%%%%%%%%%%%%%%%%%%%%%%%%%%%%%%%%%%%%%%%%%%%%%%%%%%%
The result coincides with that obtained from chiral loops,
up to exponentially small corrections which are beyond the accuracy of
L\"uscher's formula \cite{condfinitev}.
This agreement would not hold if there were any volume dependence in the
coefficients of the chiral Lagrangian, such as terms of $O(1/L^2)$.
This provides independent confirmation of the arguments of Ref.
\cite{glfiniteL}.

L\"uscher's result can be paraphrased as follows.
The pion only ``knows'' about the finite size of the box through
loop diagrams, and in particular through the pion propagating in the loop.
The vertex which emits and absorbs the pion does not itself ``know''
about the finite volume. Such knowledge would require another pion loop,
which would bring an additional exponential suppression.
This physical picture applies also to properties of the pion other than
its mass, for example $f_\pi$ and $B_K$.
It also seems reasonable to apply it to the quenched approximation.

The previous discussion has ignored the possibility of finite volume
effects due to particles being ``squeezed'' by the box.
To avoid such squeezing the box must be larger than diameter of the particle.
It is not very clear what box size this requires.
Numerical calculations of various definitions of ``wavefunctions''
show that they fall approximately exponentially in the range 0.5-1.5 fm,
with a scale of roughly $m_{\rm WF}=750$ MeV \cite{pascos,chuetal}.
This may imply a correction factor from squeezing that falls
as $\exp(-m_{\rm WF} L)$.
As long as $m_\pi < m_{\rm WF}$, the contribution from squeezing should
be suppressed.
At large enough distances the wavefunction should fall faster than an
exponential, so the effect of squeezing should be further suppressed.

Another source of finite volume dependence is loops of heavier mesons,
such as the $\rho$.
Such contributions would fall roughly as $\exp(-m_\rho L)$.
As long as $m_\pi<< m_\rho$, they should be strongly suppressed
relative to the contributions from pion loops.

I now turn to the calculation of finite volume dependence.
The loop integrals that appear are $I_1$ and $I_2$,
proportional respectively to $\int_k G$ and $\int_k G^2$.
They are related by
\begin{equation}
\label{I2fromI1}
I_2(m) = - m^2 {d \over dm^2} I_1(m) \ ,
\end{equation}
so it is sufficient to consider $I_1$.
It is simplest to work in position space
\begin{equation}
I_1(m) = {1\over f^2} \int_{|k|<\Lambda} G(k,m)
	= {1\over f^2} G_\Lambda(x\!=\!0,m) \ .
\end{equation}
The subscript $\Lambda$ indicates that the position space propagator is
smeared over a range $1/\Lambda$, due to the cut-off on the momentum integral.
At finite volume we simply substitute $G_L$ (Eq. (\ref{finiteVpion})) for $G$.
%The leading term is the infinite volume result for $I_1$ which has been
%discussed above.
We are interested in the difference
\begin{equation}
\bar{I_1}(m,L) = I_1(m,L) - I_1(m),
\end{equation}
which is given by
\begin{eqnarray}
f^2 \bar{I_1}(m,L) &=& \sum_{n_\mu\ne 0} G_\Lambda(n_\mu L_\mu,m) \\
                   &\approx& \sum_{n_\mu\ne 0} G(n_\mu L_\mu,m)
\end{eqnarray}
The smearing of the propagator has been dropped in the second line;
this is a good approximation if $L >> 1/\Lambda$, as is true in practice.
The precise form of the cut-off is irrelevant.
Using Eq. (\ref{I2fromI1}), the corresponding result for $I_2$ is
\begin{equation}
f^2 \bar{I_2}(m,L) =
\sum_{n_\mu\ne0} -m^2 {d \over d m^2} G(n_\mu L_\mu,m) \ ,
\end{equation}
In the notation of Gasser and Leutwyler these results are \cite{condfinitev}
\begin{equation}
f^2 \bar{I_1}(m,L) = g_1(m^2,L_\mu)\ ;\ \
f^2 \bar{I_2}(m,L) = m^2 g_0(m^2,L_\mu) \ .
\end{equation}
The finite volume dependence of $f_\pi$, $f_K$, $B_K$, $B_V$ and $B_A$
can now be predicted by substituting the results for $\bar{I_1}$ and
$\bar{I_2}$ into the formulae presented in previous Sections.

To actually calculate the $\bar{I_i}$,
it is convenient to use the heat-kernel representation
\begin{equation}
G(x,m) = {1\over (4\pi)^2} \int_0^\infty d\alpha
 \alpha^{-2} e^{-x^2/4\alpha} e^{-\alpha m^2} \ .
\end{equation}
Using this, the functions $g_0$ and $g_1$ turn out to be
particular cases of the general function
\begin{equation}
g_r(m,L_\mu) = {1\over (4\pi)^2} \int_0^\infty d\alpha  \alpha^{r-3}
\sum_{n_\mu\ne 0} e^{-(n_\mu L_\mu)^2/4\alpha} e^{-\alpha m^2} \ .
\end{equation}
the properties of which have been discussed extensively in
Ref. \cite{hasenfratz}.
The crucial point for present purposes is that the sum over $n_\mu$
converges rapidly, as long as $m L_\mu>>1$.
This is because each term falls like $\exp(-m n_\mu L_\mu)$, as can be seen
by evaluating the $\alpha$ integrals in saddle point approximation
\begin{eqnarray}
\label{saddleG}
G(x,m) &=& {m^2\over (4 \pi)^2} \sqrt{8\pi \over (m x)^3}\ e^{-m x}
	\left(1 + {3 \over 8 mx} + O\big({1\over (mx)^2}\big)\right) \\
\label{saddledG}
-m^2 {d \over d m^2} G(x,m) &=&
{m^2\over (4 \pi)^2} \sqrt{2\pi \over m x}\ e^{-m x}
	\left(1 - {1 \over 8 mx} + O\big({1\over (mx)^2}\big)\right) \ .
\end{eqnarray}

\begin{figure}
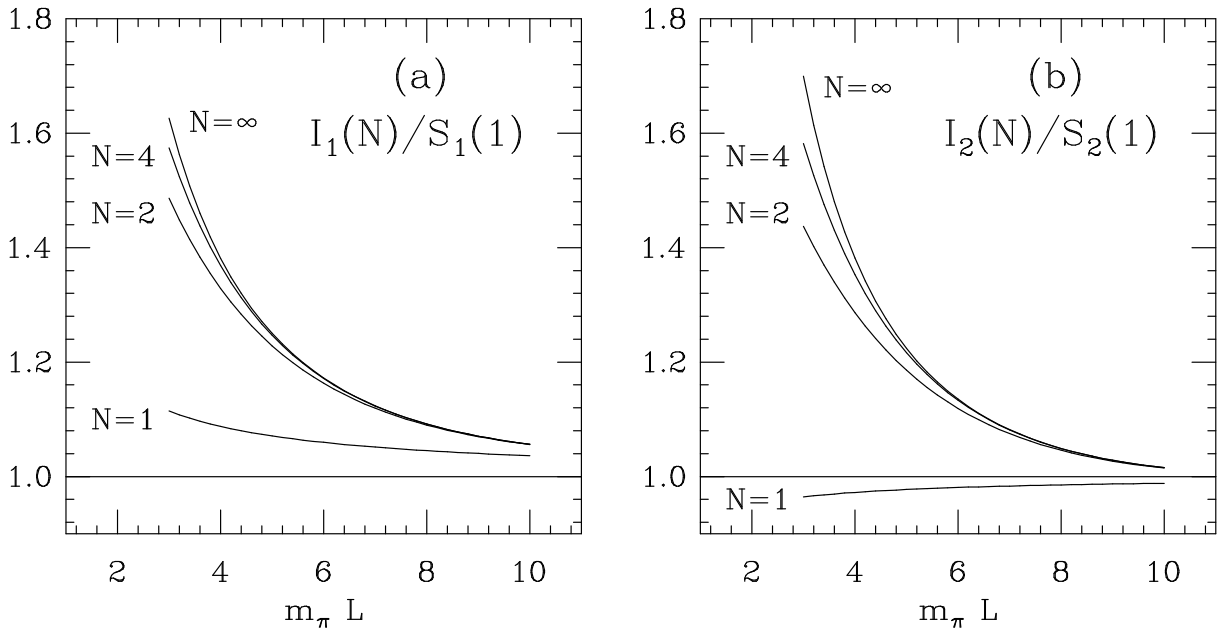

\vspace{3.5truein}
\includegraphics{fig10a.ps}
% use -1070 1450 translate
\includegraphics{fig10b.ps}
% use 530 1450 translate
\caption[ften]{\fc
The ratios of the finite volume part of the integrals $\bar{I_i}$
to the saddle point approximants $S_i$ for (a) $I_1$ and (b) $I_2$.
}
\label{ften}
\end{figure}

To illustrate the properties of the integrals I set $L_1=L_2=L_3=L$
and assume $L_4>>L$, which is the geometry used in simulations.
This means that in the sum over $n_\mu$ only the $n_i$ are non-zero.
It is useful to normalize the results by dividing by the
saddle point approximations to the terms involving propagation only
to the six adjacent periodic images, i.e. $n_i=(1,0,0)$, $(-1,0,0)$,
$(0,1,0)$ etc. These are
\begin{eqnarray}
S_1(m,L) &=& 6 {m^2\over (4 \pi f)^2} \sqrt{8\pi \over (m L)^3}\ e^{-m L} \\
S_2(m,L) &=& 6 {m^2\over (4 \pi f)^2} \sqrt{2\pi \over m L}\ e^{-m L} \ .
\end{eqnarray}
Figure \ref{ften} shows the ratios $\bar{I}/S$, computed numerically,
for different numbers of terms in the sum over $n_i$.
The notation is that $\bar{I_i}(N)$ includes the first $N$ distances,
where ``distance'' means $d=\sqrt{n_i n_i}$.
Thus for $N=1$ only the six $d=1$ terms are kept,
while for $N=4$, the twelve $d=\sqrt2$, eight $d=\sqrt3$,
and the six $d=2$ terms are also included.
Various points are noteworthy
\begin{enumerate}
\item
It is straightforward to show that the ratios depend only on the product $mL$.
\item
The results for $N=1$ show that the saddle point approximation is
accurate to better than 10\%  for $mL>3$.
If one includes the $1/mL$ corrections from Eqs. (\ref{saddleG})
and (\ref{saddledG}) then the agreement is better than 1\%.
\item
The full result ($N=\infty$) differs substantially from the
leading order term ($N=1$) at small $mL$.
Typical simulations work down to $m_\pi L=3-4$,
and for these it is clear that the leading term is not sufficient.
This was not appreciated in Refs. \cite{bkprl,sharpelat89,book}.
\item
The bulk of the difference is taken up by the $N=2$ terms,
with $N=4$ being a good approximation to the full result.
\item
The ratios for $I_1$ and $I_2$ at $N=\infty$ are very similar.
\end{enumerate}

%%% Fig 10

The need to include not only the $d=1$ term in the sum over $n_i$
raises the question of whether there are other corrections of comparable size.
Higher loop diagrams give rise to terms which are also
proportional to $\exp(-m_\pi L)$, but are suppressed by an
additional power of $m_\pi^2/16 \pi^2 f^2$.
These may be substantial for present lattice pion masses.
There are also corrections from two loop diagrams
suppressed both by a power of $m_\pi^2$,
and by an additional power of $\exp(-m_\pi L)$.
These are presumably small.

\begin{figure}
\vspace{3.5truein}
\includegraphics{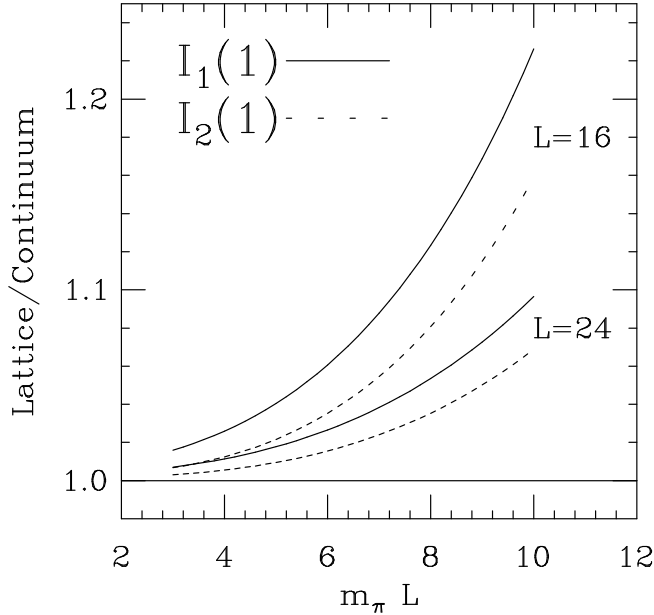}
%% use : -340 1450 translate
\caption[feleven]{\fc
Ratios of the $I_i(N\!=\!1)$ evaluated using the lattice propagator
to those evaluated using the continuum propagator.
}
\label{feleven}
\end{figure}

I close this section with a comment on the effect of the
non-zero lattice spacing, $a$.
This cuts off the momenta at $p \approx 2 \pi/a$,
which should be much larger than the chiral cut-off $\Lambda$.
This condition is satisfied in present simulations,
for which $2 \pi/a\ge 12$ GeV.
Such a large cut-off should have little effect on the propagator
over distances of $L$ or greater.
To check this, I replace the continuum propagator by that for
a lattice scalar field for which the second derivative is
discretized using the standard nearest neighbor form.
This is for illustrative purposes only: it is not clear how
the pion field should be discretized,
or indeed whether it should be discretized at all.
My choice is really the worst case one can imagine,
since discretizations involving more neighbors will give a better
approximation to the continuum derivative.
With the nearest neighbor form, the lattice propagator is
\begin{equation}
G(x,y,z,t) = \int_0^\infty d\alpha
I_x(2\alpha) I_y(2\alpha) I_z(2\alpha) I_t(2\alpha)
e^{-8\alpha} e^{-\alpha m^2} \ .
\end{equation}
here $x$, $y$, $z$ and $t$ are integers giving the distances
in lattice units, and $I_x$ is the $x$'th modified Bessel function
(and should not be confused with the integrals $I_1$ and $I_2$ under study).
The ratio of the integrals using this lattice propagator to those using
the continuum propagator are shown in Fig. \ref{feleven}.
The results are for $N=1$, but there is no significant dependence on $N$.
The ratio is close to unity, the difference not exceeding 10\%
for the values of $m_\pi$ and $L$ used in the simulations discussed
in the following section.
The effect is probably smaller than that of higher order chiral corrections.

%%% fig 11

\section{NUMERICAL PREDICTIONS AND RESULTS}
\label{snumer}
%%% sec7.tex
%% file: sec7.tex
%\section{NUMERICAL PREDICTIONS AND RESULTS}
%\label{snumer}

In this section I compare the predictions for chiral logs and
the associated volume dependence with numerical results.
For the most part the predicted volume dependence is smaller
than the statistical errors of present simulations.
My aim in such cases is to show how close we are to testing the predictions.
Only for $B_V$ and $B_A$ is there clear numerical evidence of
finite volume dependence.

I use the numerical results from quenched simulations at $\beta=6$
presented in Refs. \cite{bkprl} and \cite{hmks}.
These calculations use staggered fermions, and work
on $16^3\times40$ and $24^3\times40$ lattices.
Finite size effects can be studied by comparing results from the two lattices.
Using the lattice spacing obtained from $f_\pi$, $1/a=1.7$ GeV,
the smaller lattice has $L=1.9$ fm, the larger $L=2.8$ fm.
Both lattices are probably large enough so that the squeezing of the
pion wavefunction is a small effect.
The lattice pions have masses ranging from $m_\pi a=0.24$ to $0.41$,
or $0.41-0.70$ GeV in physical units.
The convergence of chiral perturbation theory is controlled by the ratio
$m_\pi^2/16\pi^2 f_\pi^2$, and this varies from $0.12$ to $0.36$.
Thus higher order corrections are likely to be sizable for the heavier pions.

The pions in these simulations are composed of quarks of masses
$m_q a=0.01$, $0.02$ and $0.03$.
These are combined in all possible ways, so there are results
for pions with both degenerate and non-degenerate quarks.
The predictions for quenched chiral logs given above are only for
degenerate quarks.
It turns out, however, that the results for
$m_\pi^2$, $f_\pi$, $B_K$, $B_V$ and other quantities
depend mainly on the average quark mass.
When I present results I will simply treat all pions as though
they contained two degenerate quarks each with the average mass.
%% This is really only an issue for $B_V$ and $B_A$.

\begin{figure}
\vspace{3.5truein}
\includegraphics{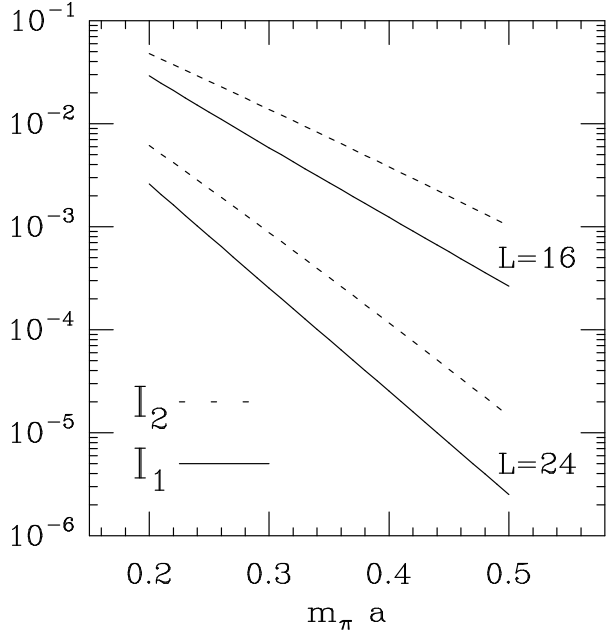}
%% use : -340 1450 translate
\caption[ftwelve]{\fc
$\bar{I_1}$ and $\bar{I_2}$, for $N=\infty$, using $1/a=1.7$ GeV.
}
\label{ftwelve}
\end{figure}

Typical chiral logarithmic corrections are numbers of $O(1)$ multiplied
by the integrals $I_1$ and $I_2$.
The finite volume dependent parts of these integrals are shown in
Fig. \ref{ftwelve}.
These are exact results, and are thus a sum of exponentials,
multiplied by powers of $m_\pi$.
Nevertheless, as the figure shows, the results are well approximated by
single exponentials for the range of quark masses considered.
The magnitude of the $\bar{I_i}$ should be compared to the
typical statistical errors in present numerical simulations,
which are rarely as small as 1\%.
For the larger lattice, and for $m_\pi a=0.24-0.41$,
$\bar{I_1}$ and $\bar{I_2}$ are always smaller than 0.4\%.
Thus, for practical purposes, $L=24$ corresponds to infinite volume.
For the smaller lattice, however, $\bar{I_2}$ reaches 3\% at $m_\pi a=0.24$,
which is a level that might be measurable.

%%% fig 12

To give an idea of the typical size of finite volume corrections
I first consider $f_\pi$ in a theory with four degenerate dynamical quarks.
This is the theory represented by a single species of staggered fermions
in the continuum limit.
{}From Eq. (\ref{fpiNflavors}) we see that
\begin{equation}
f_\pi(L) \approx f_\pi(L\!=\!\infty) (1 - 2 \bar{I_1}(m_\pi,L)) \ .
\end{equation}
It is clear from Fig. \ref{ftwelve} that to observe this effect
will require calculations with a precision of less than 1\%.
In the quenched approximation, the prediction is that there
is no finite size dependence in $f_\pi$ at leading order.
Such dependence enters only due to the diagrams of Figs. \ref{fthree}(b)
and (c), and is suppressed by $m_\pi^2/16\pi^2 f_\pi^2$.
In fact, the numerical results of Ref. \cite{hmks} show no evidence for
finite size dependence.
The errors are large, $5-10$\%, so that this does not constitute a
test of the predicted difference between full and quenched chiral logs.

I note in passing that it will be even more difficult to see the
volume dependence in $m_\pi$ calculated with dynamical quarks.
The result for $N$ degenerate flavors is \cite{condfinitev}
\begin{equation}
m_\pi^2(L) \approx m_\pi^2(L\!=\!\infty)
\left(1 + \frac1{N} \bar{I_1}(m_\pi,L)\right) \ .
\end{equation}
For the condensate, on the other hand,
 the situation is slightly better than for $f_\pi$ \cite{condfinitev}
\begin{equation}
\vev{\bar\psi\psi}(L) \approx \vev{\bar\psi\psi}(L\!=\!\infty)
\left(1 - \frac{N^2-1}{N}\bar{I_1}(m_\pi,L) \right) \ .
\end{equation}
Since the condensate can be calculated more accurately than
$f_\pi$ and $m_\pi$, this equation may offer the best chance of
testing the predictions of chiral loops in theories with dynamical quarks.
The only problem is that there is a quadratically divergent
perturbative contribution to the condensate, to which the prediction
does not apply, and which must be subtracted.

As discussed in Section \ref{seta'},
quenched chiral logs in $m_\pi^2$ come only from $\eta'$ loops.
Using Eqs. (\ref{Xfig8aresum}) and (\ref{Xfig8ares}),
the volume dependence is predicted to be
\begin{equation}
\label{mpieta'fse}
m_\pi^2(L) = m_\pi^2(L\!=\!\infty)
\exp\left({A\over N}{m_0^2\over m_\pi^2} \bar{I_2}(m_\pi,L)\right) \ ,
\end{equation}
where I have dropped terms suppressed by powers of $m_\pi^2$.
For $N=3$, the phenomenological estimate from the physical $\eta'$ mass gives
$A m_0^2/3\approx (0.5 {\rm GeV})^2$.
Thus the factor multiplying $I_2$ ranges from 1.6 to 0.55
for $m_\pi=0.41-0.70$ GeV.
This means that, although the finite volume correction
is parametrically enhanced by $1/m_\pi^2$,
and is also increased by the exponentiation,
the actual numerical enhancement is small.
The volume dependence is somewhat larger than that for the full theory,
because it is the integral $I_2$ which appears, rather than $I_1$.
The result is that, at $m_\pi a=0.24$,
the correction increases $m_\pi$ (not its square) by 2.4\% for $L=16$,
and by 0.2\% for $L=24$.
The correction decreases rapidly with increasing pion mass.
The numerical results for $m_\pi$ show no significant finite size dependence,
but the errors are about 2\%,
i.e. at same level of the expected effect \cite{hmks}.
Future quenched calculations may be able to test this prediction.

The form of the volume dependence of the condensate in the quenched
approximation is the same as that for $m_\pi^2$,
because $m_\pi^2 f_\pi^2 \propto m \vev{\bar\psi\psi}$.
Again the numerical results show no significant finite volume dependence,
but the errors are too large (5-10\% for the value extrapolated to $m_q=0$)
for this to represent a disagreement with the prediction.

\begin{figure}
\vspace{3.5truein}
\includegraphics{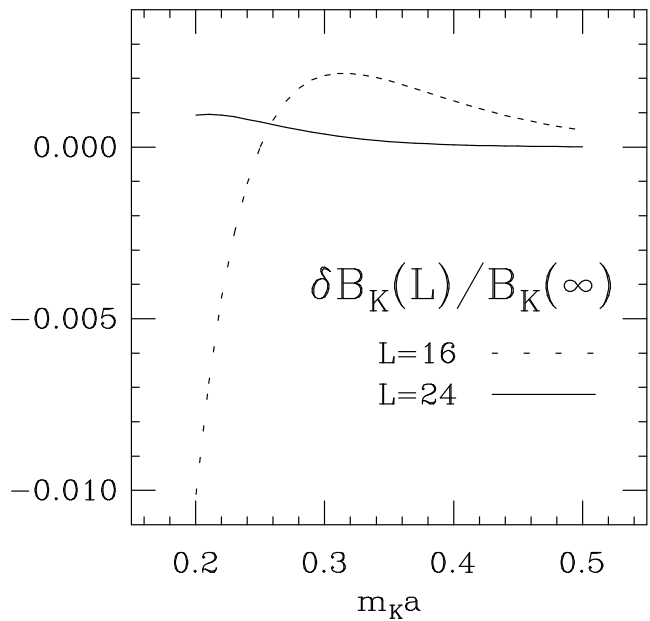}
%% use : -340 1450 translate
\caption[fthirteen]{\fc
Predictions for the fractional correction $\delta B_K(L)/B_K(L\!=\!\infty)$.
}
\label{fthirteen}
\end{figure}

I now turn to the kaon B-parameters.
For degenerate quarks, the predicted volume dependence has the same
functional form in full and quenched theories.
{}From Eq. (\ref{bkquenched}) I find
\begin{equation}
\delta B_K(L) = B_K(L) - B_K(L\!=\!\infty)
\approx  B_K(L\!=\!\infty) \left(\bar{I_2}(m_K,L) -
 2\bar{I_1}(m_K,L)\right) \ .
\end{equation}
I use $m_K$ rather than $m_\pi$ to be consistent with the name of the
B-parameter; this does not imply that the quarks in the ``kaon'' have different
masses.
In Fig. \ref{fthirteen} I show the fractional size of this correction.
There is a large cancellation between $\bar{I_2}$ and $\bar{I_1}$
in the mass range of interest, and the correction does not exceed 0.2\%.
This is unfortunate as $B_K$ can be calculated with errors as small as 1\%.
The prediction is thus that there be no finite volume dependence,
and the results of Ref. \cite{bkprl} are consistent with this.

%%% fig 13

As discussed in Section \ref{sbk},
 the chiral logs in $B_V$ and $B_A$ are enhanced
by $1/m_K^2$ over those in $B_K$, and one might hope that the
associated finite size effects can be seen.
{}From Eqs. (\ref{bv1result}) and (\ref{bv2result}), I find
\begin{equation}
\label{bvresult}
B_V = 	\frac12 B_K
    	-\frac34 (\beta_1+\beta_2) {\mu^2 \over m_K^2} I_2(m_K)
	-\frac38 (\delta_1+\delta_2)
	-\frac38 (\gamma_1+\gamma_2) I_2(m_K) + O(m_K^2)\ ,
\end{equation}
where $B_K$ is one-loop corrected result, Eq. (\ref{bkquenched}).
The enhanced chiral log is that proportional to $I_2/m_K^2$.
The important features of this prediction are:
\begin{itemize}
\item
Since $I_2/m_K^2 \propto \ln(\Lambda/m_K)$, $B_V$ should diverge
logarithmically in the chiral limit.
This is unlike $B_K$, for which the chiral logs, being proportional
to $m_K^2\ln(m_K)$, vanish in the chiral limit.
\item
The coefficient of this divergence is not related to $B_K$;
it is a new constant proportional to $\beta_1+\beta_2$.
In vacuum saturation approximation $\beta_1+\beta_2=4/3$.
Typically vacuum saturation gives the correct sign and order of
magnitude; if so, $B_V$ is predicted to diverge negatively.
\item
The extent of the enhancement is much greater than that in the
correction to the quenched $m_\pi^2$, Eq. (\ref{mpieta'fse}).
The constant $\mu=m_K^2/(m_s\!+\!m_d)$ is about 5 GeV on the lattice,
roughly ten times the corresponding quantity $\sqrt{Am_0^2/N}$.
\item
Since $I_2$ increases with $L$, $B_V$ is predicted to become
more negative as $L$ decreases.
\end{itemize}
Since $B_A=B_K-B_V$, the enhanced log in $B_A$ differs only
in sign from that in $B_V$. Thus $B_A$ is predicted to have a
positive logarithmic divergence, and to increase as $L$ decreases.
Since the numerical results for $B_K$ behave as expected,
i.e. they have a smooth chiral limit and show no volume dependence,
and since the errors in $B_K$ are much smaller than those in $B_V$ and $B_A$,
there is no additional information to be gained from studying $B_A$.
Thus I concentrate on $B_V$ in the following.

\begin{figure}
\vspace{3.5truein}
\includegraphics{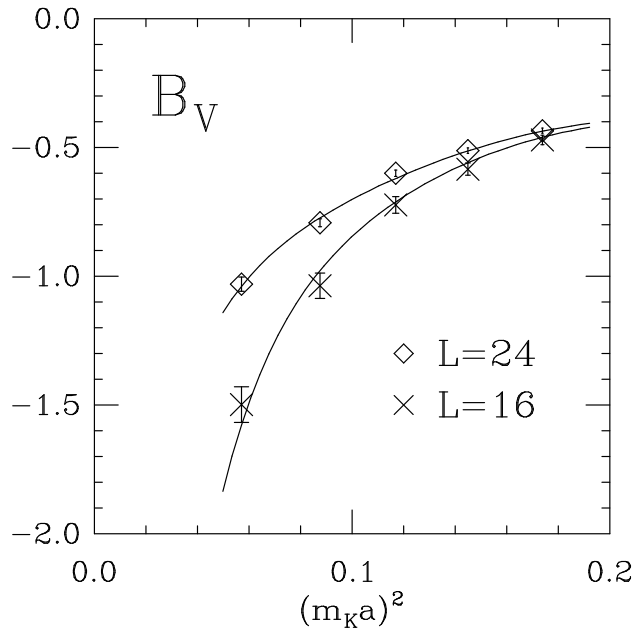}
%% use : -340 1450 translate
\caption[ffourteen]{\fc
Results for $B_V$ versus $m_\pi^2$ in lattice units. The fits are
described in the text.
}
\label{ffourteen}
\end{figure}

The numerical results of Ref. \cite{bkprl}
for $B_V$ are shown in Fig. \ref{ffourteen}.
There are two striking features:
\begin{itemize}
\item
The results appear to be diverging as the chiral limit is approached,
with the sign expected from the chiral logarithm.
\item
There is a significant finite volume dependence,
which also has the expected sign.
\end{itemize}
It was, in fact, these features that sparked the present study.

%%% fig 14

To provide a quantitative test of Eq. (\ref{bvresult}) I fit the data using
\begin{eqnarray}
\nonumber
B_V(L) &=& -{\mu^2 \over 16 \pi^2 f^2} \left(
c_1 \ln(1+1/x) + c_2 {1 \over 1+x} +c_3 + c_4 {x^2\over 1+x^2} \right) \\
\label{bvfiniteL}
&& - c_1 {\mu^2 \bar{I_2}(m_K,L) \over m_K^2} \ ,
\end{eqnarray}
where $x= m_K^2/\Lambda^2$,
$\Lambda$ being the cut-off on chiral loop integrals.
This function contains the chiral logarithm and
its associated finite volume dependence,
together with a power series in $m_K^2$.
The power series is supposed to represent all terms in Eq. (\ref{bvresult})
aside from the enhanced chiral logarithm.
This includes the normal chiral logs proportional to $m_K^2\ln(m_K)$,
which can be well approximated by a polynomial
for the range of masses under consideration.
I choose the particular form of the first two terms in Eq. (\ref{bvfiniteL})
because the loop integral is
\begin{equation}
16 \pi^2 \int_{|k|<\Lambda} G(k,m)^2 = \ln(1+1/x) - 1/(1+x) \ .
\end{equation}
The denominator of the $c_4$ term is chosen so that the function is smooth
for $x\sim1$.
Finally, the flavor symmetry breaking for staggered fermions is
accounted for by applying the prescription (\ref{staggprescription}),
with $R_{NG}=1$, to all terms in Eq. (\ref{bvfiniteL}).

Figure \ref{ffourteen} shows the best fit.
I have taken $\Lambda$ to be 0.8 GeV; the precise value is unimportant since,
to good accuracy, changes can absorbed into the other constants.
These constants are
\begin{equation}
\nonumber
c_1=0.29,\ \ c_2=-1.25,\ \ c_3=0.63,\ \ c_4=-0.37.
\end{equation}
The fit is reasonable, with a $\chi^2$ of 1.4 per degree of freedom.
I therefore conclude that the numerical results are consistent with the
expected chiral logarithm. The coefficient $c_1$ is smaller than
the result expected from vacuum saturation approximation, $c_1=1$,
but I do not consider this to be serious problem for the following reasons.
\begin{itemize}
\item
Vacuum saturation can be a poor approximation. For example,
the matrix $\cm_K$ may be as much as 50\% smaller than
the prediction of vacuum saturation \cite{sharpelat91}.
What matters here, however, are the matrix elements
of a pseudoscalar-pseudoscalar operator (cf. Eq. (\ref{lrtrans})),
and these are likely to be better approximated by vacuum saturation
\cite{bernardsonilat88}.
\item
I have assumed (by setting $R_{NG}=1$) that the vertices of
the non-Goldstone and Goldstone pions are the same,
whereas in fact the former could be smaller.
The matrix elements of some non-Goldstone bilinears are smaller by
as much as 25\% (Ref. \cite{pipinew}).
Since the loop diagrams involve the matrix element of operators
consisting of two bilinears, and also a four pion scattering amplitude,
a more substantial reduction is possible for $c_1$.
\item
There could be substantial higher order chiral perturbation theory
corrections to the coefficient of the enhanced logarithm.
That higher order terms can be substantial is illustrated by the fact
that $f_K^2/f^2=1.6$ for the heaviest lattice kaon.
\end{itemize}
The fit does not, however, demonstrate the existence of the chiral
logarithm. In particular, the coefficient $c_1$ is determined not
by the nature of the divergence, but by the finite size dependence.
If one allows the coefficient of the logarithm to differ from that
of the $\bar{I_2}$ term,
the fit favors much smaller values for the former.
In fact, a fit as good as that displayed above is obtained with no
logarithmic term at all.
This flexibility is possible because $x$ is not small, and varies over
a small range, $0.5-0.8$,
so the non-logarithmic terms in the fit function can mock up a logarithm.

\begin{figure}
\vspace{3.5truein}
\includegraphics{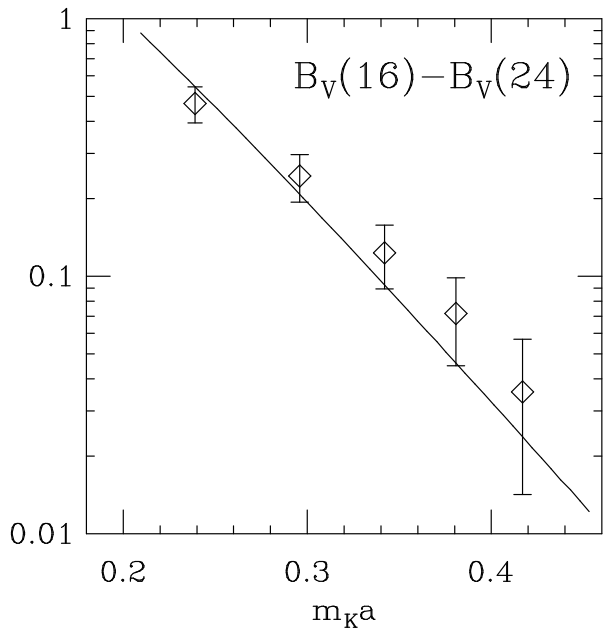}
%% use : -340 1450 translate
\caption[ffifteen]{\fc
Finite size dependence of $B_V$. The line shows the result of the
same fit as in Fig. 14.
}
\label{ffifteen}
\end{figure}

It is thus the finite volume dependence which is the strongest evidence
for quenched chiral logarithms.
The functional dependence of $B_V(16)-B_V(24)$ on $m_K$ is predicted,
leaving only the overall factor to be determined by the fit.
As shown in Fig. \ref{ffifteen}, the fit does a reasonable job
of representing the dependence on $m_K$.
The shape can be altered by higher order chiral terms, which are
proportional to $m_K^2$, so one would not expect a perfect match.
To provide a more rigorous test of the prediction
results from smaller kaon masses are needed.

%%% fig 15

\section{CONCLUSIONS}
\label{sconcl}

Chiral logarithms
can provide powerful consistency checks on numerical simulations.
The theoretical calculations are well-founded if dynamical quarks are
included. I have argued that the methods of chiral perturbation theory,
suitably modified, can also be applied in the quenched approximation.
One can imagine a program in which one first tests the results of
quenched chiral perturbation theory using simple quantities such
as $f_\pi$, $m_\pi$ and $\vev{\bar\psi\psi}$,
and then applies the predictions to other quantities
in order to test that the physics of pion loops is being properly included.
Furthermore, as I argued in the Introduction, if the calculations are
reliable one can use them to estimate the size of the errors introduced
by the quenched approximation.

Such a program requires numerical results with errors of 1\% or less,
calculated on a number of lattice sizes at a variety of pion masses.
This should be possible in the fairly near future,
as computer power approaches the TeraFlop milestone.
The numerical results presented here are encouraging, but not conclusive.
What is particularly needed is for the calculations to be pushed to smaller
pion masses. On a given lattice there is a minimal mass, $m_\pi({\rm min})$,
below which chiral symmetry is restored \cite{condfinitev}.
This minimal mass scales with size as $m_\pi({\rm min})^2 \propto L^{-3}$.
This means that $m_\pi({\rm min})L \propto 1/\sqrt{L}$,
so that, by working on larger lattices, one can push to smaller $m_\pi L$
and thereby increase the finite size effects.

A program of testing finite volume dependence also requires that
quenched chiral logarithms be calculated for other quantities.
Examples for which there is hope of good numerical results in the next
few years include the matrix elements of the $\Delta S=1$ operators
responsible for kaon decays, both CP conserving and violating,
and the properties of heavy-light systems, such as $B$ and $D$ mesons.
Calculations of the finite size effects in $f_B$ and $f_D$
in full QCD have already been done \cite{goity}.
For some quantities the quenched calculations appear
to be a straightforward extension of those presented here.
For others, one must face the problem of how best to include $\eta'$ loops.
It may be that the perturbative approach of Bernard and
Golterman \cite{bernardgolterman} will be adequate.
Or it may be that a resummation, such as that presented here
the pion mass and the condensate, is required.

It does not appear to be simple to apply the methods presented here
to all quantities. In particular, in the calculation of chiral corrections
to baryon properties, it may not be possible to uniquely assign quark
diagrams to chiral perturbation theory diagrams.
This problem is under study.\footnote{
Cohen and Leinweber \cite{leinweber} have recently begun the study
of quenched chiral logarithms in baryon properties.}

\section*{ACKNOWLEDGMENTS}
I am very grateful to Claude Bernard, Heiri Leutwyler and Apoorva Patel
for many discussions and helpful comments,
Claude Bernard and Maarten Golterman for comments on the manuscript,
and to Rajan Gupta, Greg Kilcup and Apoorva Patel
for their collaboration in obtaining the numerical results.
I also thank Norman Christ and Paul Mackenzie for helpful comments.
Parts of this work were done at the 1989 Santa Fe Workshop on QCD,
at the Aspen Center for Physics,
and at the Institute for Theoretical Physics, Santa Barbara,
all of which I thank for their excellent hospitality.
Finally I owe particular thanks to Nathan Isgur, for giving me the time
and facilities necessary to finish this project.
This work is supported in part by the DOE under contract
DE-AC05-84ER40150 and grant DE-FG09-91ER40614,
and by an Alfred P. Sloan Fellowship.

%%%%%%%%%%%%%%%%%%%%%%%%
% journal and conference list
%

\def\MPA#1#2#3{{Mod. Phys. Lett.} {\bf A#1} (#2) #3}
\def\PRL#1#2#3{{Phys. Rev. Lett.} {\bf #1}, #3 (#2) }
\def\PRD#1#2#3{{Phys. Rev.} {\bf D#1}, #3 (#2)}
\def\PLB#1#2#3{{Phys. Lett.} {\bf #1B} (#2) #3}
\def\NPB#1#2#3{{Nucl. Phys.} {\bf B#1} (#2) #3}
\def\NPBPS#1#2#3{{Nucl. Phys.} {\bf B ({Proc. Suppl.}){#1}} (#2) #3}
\def\RMP#1#2#3{{Rev. Mod. Phys.} {\bf #1} (#2) #3}
\def\NC#1#2#3{{Nuovo Cimento } {\bf #1} (#2) #3}
\def\PREP#1#2#3{{Phys. Rep.} {\bf #1} (#2) #3}
\def\ZEITC#1#2#3{{Z. Phys.} {\bf C#1} (#2) #3}

\def\brookhaven{Proc. {``Lattice Gauge Theory '86''},
             Brookhaven, USA, Sept. 1986, NATO Series B: Physics Vol. 159}
\def\fermilab#1{Proc. {``LATTICE 88''},
             Fermilab, USA, Sept. 1988, \NPBPS{9}{1989}{#1}}
\def\capri#1{Proc. {``LATTICE 89''},
             Capri, Italy, Sept. 1989, \NPBPS{17}{1990}{#1}}
\def\ringberg#1{Proc. Ringberg Workshop
``Hadronic Matrix Elements and Weak Decays'', Ringberg, Germany,
(4/88), \NPBPS{7A}{1989}{#1}}
\def\talla#1{Proc. {``LATTICE 90''},
             Tallahassee, USA, Oct. 1990, \NPBPS{20}{1991}{#1}}
\def\tsukuba{Proc. {``LATTICE 91''},
             Tsukuba, Japan, Nov. 1991, \NPBPS{}{1992}{}, to appear}

\end{document}